\documentclass{aa}
\usepackage{amsmath}
\usepackage{float}
\usepackage{caption}
\usepackage{subcaption}
\usepackage{ulem}
\usepackage{graphicx}
\usepackage{txfonts}
\usepackage{color}
\begin{document} %

    \title{Atomic line radiative transfer with MCFOST}
   \subtitle{ I. Code description and benchmarking}
   \titlerunning{atomic line mcfost}

   \author{{\sc B.~Tessore}\inst{1}, {\sc C.~Pinte}\inst{2,1}, {\sc J.~ Bouvier}\inst{1}, and  {\sc F.~M\'enard}\inst{1}}
   \institute{Université Grenoble Alpes, CNRS, IPAG, 38000 Grenoble, France 
   \and 
   School of Physics and Astronomy, Monash University, VIC 3800, Australia
   }

   \date{Received 16/10/2020; accepted 19/12/2020}


  \abstract
   {}
   {We present MCFOST-art, 
   a new non-local thermodynamic equilibrium radiative transfer solver for multilevel atomic systems. The code is embedded in the 3D radiative transfer code MCFOST and is compatible with most of the MCFOST modules. The code is versatile and designed to model the close environment of stars in 3D.}
   {The code solves for the statistical equilibrium and radiative transfer equations using the Multilevel Accelerated Lambda Iteration (MALI) method. We tested MCFOST-art on spherically symmetric models of stellar photospheres as well as on a standard model of the solar atmosphere. We computed atomic level populations and outgoing fluxes and compared these values with the results of the TURBOspectrum and RH codes. Calculations including expansion and rotation of the atmosphere were also performed. We tested both the pure local thermodynamic equilibrium and the out-of-equilibrium problems.}
   {In all cases, the results from all codes agree within a few percent at all wavelengths and reach the sub-percent level between RH and MCFOST-art. We still note a few marginal discrepancies between MCFOST-art and TURBOspectrum as a result of different treatments of background opacities at some critical wavelength ranges.
   }
   {}


    \maketitle
   
\newcommand\lprime{\ell^{\prime}}
\newcommand\lprimeprime{\ell^{\prime \prime}}
\renewcommand\l{\ell}

\section{Introduction}

Radiation processes are critical in many aspects of the evolution of astrophysical objects such as stellar atmospheres, magnetospheres and discs. Electromagnetic radiation carries the history of the emitting object to the observer at which point the object is analysed.
Through spectroscopy we can deduce fundamental parameters, abundances and unveil the complex atmospheric dynamics of many stellar objects, often via comparison with numerical models.

These numerical models require the solution of the radiative transport equation in complex and different plasma conditions.
In many astrophysical applications matter and radiation are coupled
: This is the so-called non-local thermodynamic equilibrium (non-LTE) line formation problem. 
At non-LTE, the atomic level populations are computed from a set of statistical equilibrium equations where radiation plays the central role. In turn, knowledge of emission and absorption processes, which depends on the level populations, is needed to solve for radiation.
Therefore, the radiative transport equation must be solved simultaneously with statistical equilibrium equations. This makes the solution of the non-LTE problem challenging.
Identifying the best solution to that problem depends mainly on the nature of the simulated plasma and many types of solutions exist \citep[e.g.][]{Zadelhoff2002}. 

In this paper, we present MCFOST-art, a code for the solution of the non-LTE problem for multilevel atomic systems. The code is embedded in the 3D radiative transfer code MCFOST \citep{Pinte2006,Pinte2009} and can be applied to a wide range of astrophysical problems in different geometries. 
Our main motivation for developing such a code is to address the line formation in the close environment of young stellar objects (YSO).
In particular, the evolution of classical T Tauri stars depends on the interaction between the young star and its accretion disc, on distances of a few stellar radii.
This complex interaction leads to stellar winds and magnetospheric accretion and ejection processes. These processes have a strong impact on spectral lines and disentangling their typical radiative signatures is a challenging task.

Previous modelling of the environment of T Tauri stars, their magnetosphere \citep{Hartmann94, Muzerolle2001}, stellar wind, and disc wind regions \citep{Lima2010, Kurosawa2011} show the difficulty to account for all observational signatures of these stars, from their variability to the shape of emission lines. 
Even a detailed modelling of a specific T~Tauri star, with the most updated magneto-hydrodynamics (MHD) models, results in a marginal agreement between synthetic observables and observations \citep{Alencar2012}. 

Most of magnetospheric emission lines synthesis used the Sobolev approximation \citep{Sobolev57,Rybicki78} to solve the radiative transfer equation. 
This approximation is applicable when large velocity gradients (i.e. large compared to the intrinsic width of atomic lines) are encountered in a plasma. It greatly simplifies the solution of the transfer equation. However, in the case of magnetospheres, the required conditions for the Sobolev approximation to hold are not always met, ultimately impacting the level populations. 
In our code, we use an accelerated $\Lambda$-iteration method \citep{ali_fundation} with a preconditioning of the statistical equilibrium equations \citep{RybickiHummer1991} to solve the coupled radiative transfer and statistical equilibrium equations. Unlike the Sobolev approximation, our approach does not make any assumptions about the local conditions in the plasma.

We tested MCFOST-art on 1D spherically symmetric models including a range of thermodynamic quantities to capture a variety of physical conditions existing in stellar plasma.
In \S \ref{physical_model} we give the details of the physical problem and how it is solved. In \S \ref{impl_mcfost} we describe the implementation of our code in MCFOST, and in \S \ref{benchmarks} we test the accuracy and precision of our code against benchmark cases. Finally, we give our conclusions in \S \ref{conclusion}.

\section{Non-NLTE radiative transfer problem}\label{physical_model}

In multilevel non-LTE problems, the main difficulty is to find a self-consistent solution of the equations for the atomic level populations with a solution of the transport equation for radiation. In this section we describe the coupling between these equations.
\subsection{Radiative transfer equation}\label{sect:RTE}

  The unpolarised radiative transfer equation along a ray of length $ds$ in the direction $\mathbf{n}$ is written as
   \begin{equation}\label{eq:RTE}
      \dfrac{dI ({\nu},\mathbf{n})}{ds} = -\chi ({\nu},\mathbf{n}) I ({\nu},\mathbf{n}) + \eta ({\nu},\mathbf{n})
   \end{equation}
where, $I ({\nu},\mathbf{n})$ is the specific intensity at a frequency $\nu$ in the direction $\mathbf{n}$, $\chi$ is the total (line and continuum) opacity (true absorption and scattering $\sigma$), and $\eta$ is the emissivity.

We use the formulation of \cite{RybickiHummer92b} to express the radiative transfer equation, and we define the radiative transfer coefficients $U_{\l \lprime}$ and $V_{\l \lprime}$ as

\begin{equation}\label{eq:RTCl}
\left.
    \begin{array}{l}
    U_{\l \lprime}(\nu, \mathbf{n})  = \dfrac{h\nu}{4\pi}  A_{\l \lprime} \psi (\nu, \mathbf{n}) \text{~if~$\l>\lprime$} \\\\
    V_{\l \lprime}(\nu, \mathbf{n})  = \dfrac{h\nu}{4\pi}  B_{\l \lprime} \phi (\nu, \mathbf{n})
    \end{array}
\right.
\text{,}
\end{equation}
where $A_{\l \lprime}$ and $B_{\l \lprime}$ are the Einstein's coefficients and $\phi (\nu, \mathbf{n})$ and $\psi (\nu, \mathbf{n})$ the absorption and the emission profiles, respectively, for the transition $i \to j$\footnotemark. 
We further assume complete frequency redistribution, implying an equivalence between the absorption coefficient appearing in $V_{\l \lprime}$  and the emission coefficient appearing in $U_{\l \lprime}$, that is $\psi (\nu, \mathbf{n})= \phi(\nu, \mathbf{n})$. 

For a continuum transition the radiative transfer coefficients are given by

\footnotetext{We adopt the following convention: $j$ and $i$ refer to the upper and lower levels of a transition, respectively. For atomic levels ordered by increasing energies, $E_{\ell}$, this implies $E_{j} > E_{i}$.
In other cases, when the ordering of levels is unimportant, we use the indexes $\l,\,\lprime,\, \lprimeprime$...}

\begin{equation}\label{eq:RTCc}
\left.
    \begin{array}{l}
    U_{\l \lprime}  = \alpha_{\l \lprime} (\nu) (\dfrac{n_{\lprime}^\ast}{n_\l^\ast}) \dfrac{2h\nu^3}{c^2}e^{-h\nu/kT}  \text{~if~$\l>\lprime$} \\\\
    V_{\l \lprime}  = \alpha_{\l \lprime} (\nu) (\dfrac{n_{\lprime}^\ast}{n_\l^\ast})e^{-h\nu/kT} \text{~if~$\l>\lprime$}\\\\
    V_{\l \lprime}  = \alpha_{\l \lprime} \text{~if~$\l<\lprime$}
    \end{array}
\right.
\text{,}
\end{equation}
where $\alpha(\nu)$ is the photoionisation cross-section, $T$ the gas temperature, and $n_{\l}^\ast$ the population (number density) of level $\l$, evaluated at local thermodynamic equilibrium (LTE).

Finally, the emissivity $\eta_{ij}$ and the opacity $\chi_{ij}$ for a transition between an upper level $j$ and a lower level $i$ are written as

\begin{equation}\label{eq:opacity}
\left.
    \begin{array}{l}
    \chi_{ij}(\nu, \mathbf{n})  = n_i V_{ij}(\nu, \mathbf{n}) - n_j V_{ij}(\nu, \mathbf{n})\\
    \eta_{ij}(\nu, \mathbf{n}) = n_j U_{ji}(\nu, \mathbf{n})
    \end{array}
\right.
\text{,}
\end{equation}
where $n_i$ and $n_j$ are the lower and upper level populations, respectively. The total opacity and emissivity for all transitions of all atoms and other sources of opacity add up to the total opacity $ \chi(\nu, \mathbf{n})$ and emissivity $ \eta(\nu, \mathbf{n})$,
\begin{equation}\label{eq:opacityTot}
\left.
    \begin{array}{l}
    \chi(\nu, \mathbf{n})  = \displaystyle\sum \limits_{j, i<j} n_i V_{ij}(\nu, \mathbf{n}) - n_j V_{ij}(\nu, \mathbf{n}) + \chi_c\\
    \eta(\nu, \mathbf{n}) = \displaystyle\sum \limits_{j, i<j} n_j U_{ji}(\nu, \mathbf{n}) + \eta_c
    \end{array}
\right.
\text{,}
\end{equation}
where $\eta_c$ and $\chi_c$ are the sources of continuous emissivity and opacity evaluated at LTE (the background opacity, see \S \ref{backgr_opac}).

The solution of Eq. \ref{eq:RTE} along a ray of length $ds$ is straight forward if $\chi$, $\eta$, the populations $n$, and  the properties of the atmosphere (temperature, velocity fields, etc...) are known.
Defining the optical depth $\tau \coloneqq \tau(\nu, \mathbf{n}; s)$ as $d\tau (\nu, \mathbf{n}) = -\chi(\nu, \mathbf{n}; s)\,ds$ 
it reads:

\begin{equation}\label{eq:RTEsol}
    I(\nu, \mathbf{n}; \tau_{2}) = I(\nu, \mathbf{n}; \tau_{1}) e{^{-(\tau_2 - \tau_1)}} +  \displaystyle\int\limits_{\tau_1}^{\tau_2} S(\nu, \mathbf{n};\tau)\,e{^{-(\tau-\tau_1)}} d\tau \text{,}
\end{equation}

where $S(\nu, \mathbf{n}; \tau) = \dfrac{\eta (\nu, \mathbf{n}; \tau)}{\chi (\nu, \mathbf{n}; \tau)}$ is the source function at optical depth $\tau$, at the frequency $\nu$ in the direction $\mathbf{n}$.

In general, the emissivity and opacity depend on the level populations which in turn depend on the intensity. Therefore, a self-consistent solution of the radiative transfer equation with a solution of statistical equilibrium equations is required, the so-called non-LTE problem.

\subsection{$\Lambda -$iteration}
The solution of Eq. \ref{eq:RTE} can be recast as
\begin{equation}\label{eq:Lambda1}
     I(\nu, \mathbf{n}) = \Psi(\nu, \mathbf{n}) [\eta (\nu, \mathbf{n})]\text{,}
\end{equation}
where $\Psi$\footnotemark is a matrix operator whose elements depend on the level populations \citep[for more details see e.g. ][hereafter, HM14]{HubenyMihalas2014}.

\footnotetext{It is common in radiative transfer problems to see the $\Lambda$ operator when it comes to "$\Lambda -$iteration". The relation between the $\Psi$ and $\Lambda$ operators is given in \cite{RybickiHummer1991} as $\Psi = \Lambda / \chi$.}

This equation gives the solution of the intensity if all elements of the $\Psi$ operator, which are functions of the atomic populations, are known. However, in practice, building the full $\Psi$ operator is never affordable and the emissivity appearing in Eq. \ref{eq:Lambda1} is substituted by an old value $\eta^{\dagger}$ evaluated from a known estimate of the level populations (i.e. old values obtained with a previous iteration). The new intensity obtained by the application of the operator on the old emissivity is then used to compute a new value of the level populations and of the emissivity, which are used to compute a new value of the intensity. This iterative process, between old and new values of the populations, is known as $\Lambda-$iteration and it is repeated until convergence.
Although $\Lambda-$iteration is very efficient in an optically thin region, it suffers convergence problems in very optically thick regions. This drawback of classical $\Lambda-$iteration can be overcome by a slight modification of Eq. \ref{eq:Lambda1} \citep[see][]{Cannon73a,Cannon73b,Scharmer81,ali_fundation}, that is recasting Eq. \ref{eq:Lambda1} in 
\begin{equation}\label{eq:Lambda2}
     I(\nu, \mathbf{n}) = (\Psi(\nu, \mathbf{n}) - \Psi^{\ast}(\nu, \mathbf{n})) [\eta^{\dagger} (\nu, \mathbf{n})] + 
      \Psi^{\ast}(\nu, \mathbf{n}) [\eta (\nu, \mathbf{n})] \text{,}
\end{equation}
with an approximate operator $\Psi^{\ast}(\nu, \mathbf{n})$ built from a subset of the original operator. Equation \ref{eq:Lambda2} improves the convergence of the classical $\Lambda$-iteration method in optically thick regions and has been named ALI for accelerated $\Lambda$-iteration (after the work of \citet{Hamann85} and \citet{WernerHusfeld85}).

In most applications, the diagonal of the full operator is used, although other studies recommended a tri-diagonal approximate operator \citep[see for instance ][]{Hennicker2018}. Higher order approximate operators can improve the convergence of the radiative transfer problem at the price of computational time and memory storage. For 2D radiative transfer problems,
\cite{Auer94} suggested to use the diagonal part of the operator in combination with an algorithm to accelerate the convergence (see \S \ref{acceleration}).
In our code, we use the diagonal part of the full operator as a compromise between speed and accuracy.

\subsection{Statistical equilibrium equations}\label{sect:SEE}
The level populations of an atom in a stellar plasma is a function of the rates of transitions populating or depopulating a given level. These rates in turn are a function of level populations, collisions, and radiation.

The statistical equilibrium equations (SEE) for level $\l$ is given by\footnotemark
\begin{equation}\label{eq:SEE}
    \displaystyle\sum \limits_{\lprime}^{} n_\lprime (C_{\lprime \l} + R_{\lprime \l}) = 
    n_\l \displaystyle\sum \limits_{\lprimeprime}^{} (C_{\l \lprimeprime} + R_{\l \lprimeprime})
    \text{,}
\end{equation}
where $n_{\l}$ is the population of level $\l$, and $C_{\lprime \l}$, and $R_{\lprime \l}$ the collisional and radiative rates for a transition between level $\lprime$ and $\l$, respectively.

\footnotetext{We neglect both the advective and non-stationary terms as they are in general negligible for non-relativistic flows.}

\subsubsection{Radiative rates}
Unlike collisional rates, radiative rates are a product of the radiative transfer code.
These rates can be expressed in terms of the radiative transfer coefficients and are written as
\begin{equation}\label{eq:RR}
    R_{\l \lprime}= \displaystyle\int d\Omega \displaystyle\int \dfrac{d\nu}{h\nu} \{ U_{\l \lprime}(\nu,\mathbf{n}) + V_{\l \lprime}(\nu,\mathbf{n}) I(\nu,\mathbf{n})\} \text{,}
\end{equation}
where the integration is carried out over frequency $\nu$ and all solid angles $\Omega$.
For line transitions these rates are written as

\begin{equation}\label{eq:RRl}
\left.
    \begin{array}{l}
    R_{ij}= B_{ij} \mathcal{\overline{J}} \\
    R_{ji}= A_{ji} + B_{ji} \overline{\mathcal{J}} \\
    \end{array}
\right.
\text{,}
\end{equation}
where $B_{ij}$, $B_{ji}$, and $A_{ji}$ are the Einstein's coefficients, and  $\overline{\mathcal{J}} $ the mean intensity integrated over the line given by
\begin{equation}\label{eq:Jbar}
    \overline{\mathcal{J}} =  \displaystyle\int \dfrac{d\Omega}{4\pi} \displaystyle\int d\nu\, I(\nu, \mathbf{n})\phi (\nu, \mathbf{n}) \text{,}
\end{equation}

where $\phi (\nu, \mathbf{n})$ is the line absorption profile, also appearing in Eq. \ref{eq:RTCl}.
For continuum transitions they are as follows:
\begin{equation}\label{eq:RRc}
\left.
    \begin{array}{l}
    R_{ij}= \dfrac{4\pi}{h} \displaystyle\int d\nu\, \alpha_{ij} (\nu) \mathcal{J}(\nu)\\
    R_{ji}=  \dfrac{4\pi}{h}\dfrac{n_{i}^{\ast}}{n_{j}^{\ast}} \displaystyle\int d\nu\, \alpha_{ij}(\nu) e^{-h\nu/kT} \{\dfrac{2h\nu^3}{c^2} + \mathcal{J}(\nu)\}\\
    \end{array}
\right.
\text{,}
\end{equation}
where $\mathcal{J}(\nu)$ is the mean radiation field given by
\begin{equation}\label{eq:Jmean}
    \mathcal{J}(\nu) = \displaystyle\int \dfrac{d\Omega}{4\pi} \, I(\nu, \mathbf{n})
\end{equation}

\subsubsection{Matrix form of the SEE}
Starting from Eq. \ref{eq:SEE}, passing the right-hand side term on the left-hand side and noticing that $n_l = \displaystyle\sum \limits_{\lprime}^{} n_\lprime \delta_{\l \lprime}$ with $\delta_{\l \lprime}$ non-zero only for $\lprime=\l$, we can factorise out $\displaystyle\sum \limits_{\lprime}^{} n_\lprime$ and write the SEE as
\begin{equation}\label{eq:SEE2}
    \displaystyle\sum \limits_{\lprime}^{} n_\lprime \Gamma_{\lprime \l} = 0 \text{,}
\end{equation}
where the rate matrix $\Gamma_{\lprime \l}$ is
\begin{equation}\label{eq:Gamma}
    \Gamma_{\lprime \l} = R_{\lprime \l} + C_{\lprime \l} - \delta_{\l \lprime} \displaystyle\sum \limits_{\lprimeprime}^{}
     \{ R_{\l \lprimeprime} + C_{\l \lprimeprime} \}
\end{equation}

\subsection{MALI method}
Multilevel radiative transfer is a complex problem to solve because of the dependence of the radiation field on the populations, and of the populations on the radiation field.
Using Eqs.\ref{eq:Lambda2}  and \ref{eq:RR} the rate matrix in Eq. \ref{eq:Gamma} for a multilevel atom reads:

\begin{equation}\label{eq:Gamma1}
\begin{split}
    \Gamma_{\lprime \l} = C_{\lprime \l} + \displaystyle\int d\Omega  \displaystyle\int \dfrac{d\nu}{h\nu} \{ U_{\lprime \l} + V_{\lprime \l}I_{\rm eff}  (\nu, \mathbf{n}) \} \\
     - \delta_{\l \lprime} \displaystyle\sum \limits_{\lprimeprime}^{} [
     C_{\l \lprimeprime} + \displaystyle\int d\Omega \displaystyle\int \dfrac{d\nu}{h\nu}\{  U_{\l \lprimeprime} + V_{\l \lprimeprime}I_{\rm eff}  (\nu, \mathbf{n})\}] \\
     - \displaystyle\int d\Omega \displaystyle\int \dfrac{d\nu}{h\nu} \displaystyle\sum\limits_{j>l}^{} (n_{l}V_{lj} - n_{j} V_{jl} ) \displaystyle\sum\limits_{i<\lprime}^{} \Psi^{\ast}(\nu, \mathbf{n})U_{\lprime i} \text{,}
\end{split}
\end{equation}
where $I_{\rm eff}  (\nu, \mathbf{n}) = I^{\dagger}(\nu, \mathbf{n}) - \Psi^{\ast}(\nu, \mathbf{n}) \eta^{\dagger}(\nu, \mathbf{n}) $ is an effective radiation field built with known quantities (old values). In Eq. \ref{eq:Gamma1} we assumed that background opacities are constant within subsequent iterations. 

The system of equations \ref{eq:SEE2} is nonlinear in the new populations as it involves the product of the form $ \propto n_{\l} \times n_{\lprime}$ in Eq. \ref{eq:Gamma1}.
The multilevel accelerated lambda iteration method (hereafter, MALI) proposed by \cite{RybickiHummer1991}, latter improved by \cite{Uitenbroek2001}, uses the operator splitting method and a full preconditioning to make Eqs. \ref{eq:SEE2} and \ref{eq:Gamma1} linear in the new populations. The coupled equations of statistical equilibrium with the radiation field are finally written as follows:

\begin{equation}\label{eq:Gamma2}
\begin{split}
   \displaystyle\sum \limits_{\lprime}^{} n_\lprime  C_{\lprime \l} + \displaystyle\sum \limits_{\lprime}^{} n_\lprime \displaystyle\int d\Omega \displaystyle\int \dfrac{d\nu}{h\nu} \{ U_{\lprime \l}^{\dagger} + V_{\lprime \l}^{\dagger}I_{\rm eff}  (\nu, \mathbf{n}) \} \\
     - \displaystyle\sum \limits_{\lprime}^{} n_\lprime \delta_{\l \lprime} \displaystyle\sum \limits_{\lprimeprime}^{} [
     C_{\l \lprimeprime} + \displaystyle\int d\Omega \displaystyle\int \dfrac{d\nu}{h\nu}\{  U_{\l \lprimeprime}^{\dagger} + V_{\l \lprimeprime}^{\dagger}I_{\rm eff}  (\nu, \mathbf{n})\}] \\
     - \displaystyle\sum \limits_{\lprime}^{} n_\lprime \displaystyle\int d\Omega \displaystyle\int \dfrac{d\nu}{h\nu} \displaystyle\sum\limits_{j}^{} (n_{l}^{\dagger}V_{lj}^{\dagger} - n_{j}^{\dagger} V_{jl}^{\dagger} ) \displaystyle\sum\limits_{i<\lprime}^{} \Psi^{\ast}(\nu, \mathbf{n})U_{\lprime i}^{\dagger} = 0
\end{split}
\end{equation}%

The last term in Eq. \ref{eq:Gamma2}, which disappears in the classic $\Lambda$-iteration \cite[][]{RybickiHummer92b,Uitenbroek2001} is called the cross-coupling term. From Eq. \ref{eq:opacity} the term $\displaystyle\sum\limits_{j}^{} (n_{l}^{\dagger}V_{lj}^{\dagger} - n_{j}^{\dagger} V_{jl}^{\dagger} )$ represents a summation over all opacities of all transitions with a plus sign if $l < j$ or a minus sign if $l>j$, that is $\displaystyle\sum\limits_{j>l}^{} \chi_{lj}^{\dagger} - \displaystyle\sum\limits_{j<l}^{} \chi_{jl}^{\dagger}$.

\subsection{Coherent electron scattering}\label{cohes}
In some cases electron scattering is a dominant source of opacity and its effect on both the continuum and spectral lines has to be taken into account \citep{Hillier91}. However, the evaluation of the scattering emissivity involves the calculation of a scattering integral whose expression in the observer's frame is not trivial \citep{RybickiHummer94}.
We evaluate the electron scattering emissivity by setting it to the mean intensity (coherent approximation), which is in turn determined iteratively through classical $\Lambda$-iterations, as suggested by \cite{RybickiHummer92b}. This method is accurate enough for the main range of applications of the code, but becomes inaccurate for hot stellar wind where the electron density is very high and the ratio of scattering to true absorption is large.

\subsection{Level dissolution and occupation probability}

To do reliable radiative transfer simulations it is necessary to allow for line overlap and merging close to series limits. The MALI formulation of the non-LTE problem is already capable of dealing with overlapping lines, and we include the formalism of occupation probabilities of \cite{HummerMihala88} to allow for lines to merge at the series limits. 
Because of the presence of perturbers, each bound state $i$ of an atom has a probability $w(i)$, with respect to the same state of a similar isolated atom, to remain bound. Conversely, $1-w(i)$ represents the probability of state i to be dissolved, in other words, to belong to the continuum states. Dissolved states contribute to a pseudo-continuum opacity beyond the series limits. 
Our implementation of the occupation probabilities formalism is similar to that of \cite{Hubeny94}, based on the work of \cite{Dappen87}, and summarised in \cite{HillierMiller98}. We further assume, as in \cite{HillierMiller98}, that if $w(i)$ is the probability of level $i$ to be undissolved and $w(j)$ this probability for level $j$, then if $j$ is undissolved, $i$ is necessarily undissolved as it lies in a lower energy state. Therefore, all upward rates in the statistical equilibrium equations have to be multiplied by $w(j)/w(i)$, the probability that the upper level $j$ is undissolved given $i$ is undissolved. In \S \ref{star_continuum} the impact of level dissolution on stellar continua is shown.

\subsection{Acceleration of the convergence}\label{acceleration}
To speed up the MALI method we implemented the method of acceleration of \cite{Ng_acc} with general orders as defined in \cite{Auer94} and \cite{Uitenbroek2001}. This method uses $N_{sol} + 2$ previous solutions and accumulated iteration after iteration to extrapolate the new value of the solution by minimising the residual between the previous iterations. The latter step is called Ng's iteration.
When using Ng's accelerations, it is important to impose a delay before accumulating solutions, as extrapolating the new solution too early may result in negative or inconsistent values. Following \cite{Auer94}, we recommend doing some MALI iterations between each extrapolation to let the solution settle down. Therefore, after $N_{start}$ MALI iterations (i.e. without acceleration), we start accumulating $N_{sol} + 2$  solutions up to the Ng's iteration. We repeat this process every $N_{pending}$ MALI iterations. In our tests, we use $N_{sol} = 2$, $N_{start} = 5$ and $N_{pending} = 5$.
We note that setting $N_{pending}$ to 0 mainly results in extra overhead due to the matrix inversion required in the minimisation procedure \citep[see for instance][]{Uitenbroek2001}.

The main drawback of Ng's acceleration for multidimensional models is the memory required to store $N_{sol} + 2$ solutions for each atom treated in non-LTE. Following \cite{Auer94}, using a diagonal operator in addition to Ng's iterations is a good compromise between speed (of convergence) and memory requirements in multidimensional models. Whereas in 1D geometry memory storage is usually not a limitation, we discuss the practicability of Ng's acceleration scheme in multidimensional models in a future paper.

\section{Implementation in MCFOST }\label{impl_mcfost}

The formulation of the non-LTE line transfer of \cite{RybickiHummer92b}, although developed for 1D grids, can be easily adapted for multidimensional models. 
 \cite{Auer94} applied the MALI method to a 2D Cartesian grid, while \cite{HB14_3Dnlte} applied it to a 3D spherical grid. Recently, \cite{DeCeuster2020} applied the method to an unstructured grid.

MCFOST-art, which stands for MCFOST atomic radiative transitions, is an ensemble of modules implemented in MCFOST \citep{Pinte2006,Pinte2009}. A 3D radiative transfer code, MCFOST is written in modern Fortran, dedicated to the modelling of dust emission and molecular lines in the environment of young stars (e.g. protoplanetary discs, circumstellar envelopes). This code has been coupled with SPH codes \citep[e.g.][]{Price2018} and MHD codes \citep[e.g.][]{coupling_mcfost_pluto1}. Currently MCFOST handles three types of grids: a cylindrical grid dedicated to the modelling of discs, a spherical grid, and an unstructured grid based on a Voronoi tessellation \citep[see e.g.][]{Camps2013}. 
The solution of the line transfer equation is done by ray-tracing: several rays are propagated in specific directions along which Eq. \ref{eq:RTE} is solved for each wavelength simultaneously.

In our new modules, we propagated rays from each cell centre (spatial grid units in 2D and 3D geometries) in directions defined by the angular quadrature scheme (see below). 
Cells represent the smallest resolution elements of the code and all quantities are constant within them. If the model has a non-zero velocity field, the velocity is projected in the direction of propagation of the ray. When velocity fields are present, it is important to maintain a proper spatial discretisation to prevent nonphysical features due to large velocity gradients \citep{Ibgui2013}. The MCFOST code detects if the projected velocity between two grid points is too large and linearly interpolates the projected velocity accordingly. 

To produce images, we solved the intensity out of each pixel by integrating Eq. \ref{eq:RTE} along rays centred on each pixel.

The solution of Eq. \ref{eq:RTE} along a ray, specified by a vector direction $\mathbf{n}$, is written as
\begin{equation}\label{eq:RTEmcfost}
    I_{0} (\nu, \mathbf{n}) = I_{b} (\nu, \mathbf{n})\,e^{-\tau_{tot}(\nu, \mathbf{n})} + \displaystyle\sum\limits_{k}^{}
    S_{k} (\nu, \mathbf{n}) \, 
    (1 - e^{-d\tau_{k}(\nu, \mathbf{n})})\, e^{-\tau_{k}(\nu, \mathbf{n})} \text{,}
\end{equation}
where $I_{0} (\nu, \mathbf{n})$ is the emerging specific intensity in a given direction; $I_{b} (\nu, \mathbf{n})$ is the intensity at the inner boundary of the model, typically this is the stellar radiation if a ray intersects the star; $\tau_{tot}$ is the total optical depth from the boundary to the observer; $S_k$ the source function for cell k; and $\tau_{k}$ the optical depth at the inner (entrance) boundary of a cell. In Eq.~\ref{eq:RTEmcfost}, the sum represents the contribution of each cell, along the ray path, to the outgoing radiation.

The term $d\tau_k = \tau_{k+1} - \tau_{k} = \chi_k \, s$ is the optical depth accumulated by the ray as it travels a distance $s$ inside a cell of opacity $\chi_k$.
The intensity outgoing in direction $\mathbf{n}$ from a specific cell, after travelling a distance $s(k+1) - s(k)$ in the cell is written as

\begin{equation}\label{eq:RTEcell}
\begin{split}
       I_{k+1} (\nu, \mathbf{n}) = I_{k} (\nu, \mathbf{n}) \,+
   S_{k} (\nu, \mathbf{n}) \,(1 - e^{-d\tau_{k}(\nu, \mathbf{n})})\, e^{-\tau_{k}(\nu, \mathbf{n})} \text{,}
\end{split}
\end{equation}

The approximate operator $\Psi^{\ast}$ appearing in eqs. \ref{eq:Lambda2} and \ref{eq:Gamma2} is evaluated by solving the transfer equation with a source function, which is 1 at cell k and 0 elsewhere, that is $S_k = \delta(\tau - \tau_k)$. According to Eq. \ref{eq:RTEcell}, this is given by

\begin{equation}\label{eq:PsiOpk}
    \Psi^{\ast}_k (\nu, \mathbf{n}) = (1 - e^{-d\tau_{k}(\nu, \mathbf{n})})\,/\, \chi_{k}^{\dagger}(\nu, \mathbf{n}) \text{,}
\end{equation}
where the numerator is the diagonal of the $\Lambda$ operator, that is
\begin{equation}\label{eq:lambdaOpk}
    \Lambda^{\ast}_k (\nu, \mathbf{n}) = (1 - e^{-d\tau_{k}(\nu, \mathbf{n})})
\end{equation}

This satisfies the following conditions in the two extreme cases: in optically thin regions it is 0, which corresponds to classical $\Lambda$-iterations, and it approaches 1 in optically thick regions.
Although this method allows for fast integration of the radiative transfer equation, it depends on the numerical resolution, compared to other methods that use linear or cubic interpolations between grid points.

The part of the code that handles the propagation within the grid is presented in \cite{Pinte2006}. In the following, we present in some detail the solution of the non-LTE atomic line formation problem in the code.

\subsection{Initial solution}\label{initial_cond}

In MCFOST-art, there are currently two choices for the initialisation of the level populations of the non-LTE problem: populations are evaluated at LTE and populations are obtained by solving the SEE with the radiation field set to zero. 
The choice of the initial solution is the critical point of the non-LTE problem. 
If the initial guess is too far from the solution, the convergence can be slowed down or even fail.
In the eventuality that none of these choices lead to convergence, we implemented the collisional-radiative switching method of \cite{Hummer_Voels88}. This method allows for a smooth transition between a collision dominated initial solution (i.e. at LTE) to a non-LTE initial solution.
Alternatively, the populations from a previous calculation are given.

\subsection{Line profiles}
The absorption profile for each atomic bound-bound transition can either have a Gaussian or a Voigt shape. The choice of the shape of the absorption profile depends on the line and the problem.

The width of the absorption profile, $v_{D}$,  is given by the sum of the line thermal width and the microturbulence $\xi$ as follows:

\begin{equation}\label{eq:vd}
    v_D  = \sqrt{2 k_b T / m  + \xi^2} \text{,}
\end{equation}
where $k_b$ is the Boltzmann constant, $T$ the temperature, and $m$ the mass of the atomic species.

The damping parameter is the sum of the radiative damping and Van Der Waals and Stark broadenings.
The radiative damping of a line transition $j \to i$, $\Gamma_j$, is computed by summing the Einstein $A_{ji}$ coefficients for all (allowed) transitions for which $j$ is an upper level as follows:
\begin{equation}\label{eq:NaturalDamping}
    \Gamma_j = \displaystyle\sum\limits_{i<j} A_{ji}
\end{equation}
In the latter expression, we neglected the effect of stimulated emission.

Pressure broadening takes into account van der Waals and Stark (linear and quadratic) broadenings computed in the Lindholm theory. 
We computed interaction constants using the Uns\"old formula \citep{Unsold1955} and evaluated the quadratic Stark broadening following \citet[][Eq. 11.33]{Gray2008}.
Linear Stark broadening was computed following \cite{Sutton1978}.
The treatment of the Stark broadening, especially for hydrogen lines is approximate and is only accurate for the first lines of each series. In the future, we plan to add a more accurate treatment of the Stark broadening \citep[see e.g.][]{Stehle99}.

\subsection{Wavelength grid}\label{waves_grid}
The wavelength grid used for non-LTE calculations was built from individual atomic transitions. Firstly, we set up a grid of continuum transitions at a moderate resolution. This continuum grid was used to solve the continuum radiative transfer equation. 

Secondly, lines were gathered in groups depending on their bounds (i.e. overlapping regions). Line bounds range from $-V$ to $+V$, where $V$ is a multiple of the line Doppler width (Eq. \ref{eq:vd}). 
The line bounds represent the interaction zone of the line with the radiation field. Since in moving atmosphere a line feels the radiation from different regions, the line bounds include the maximum shift in frequency due to the velocity fields.
However, the local (i.e. in the co-moving frame) line profile is not affected by velocity fields.

Each group of lines was sampled with a constant resolution in kilometers per second. Once wavelength grids for each group were determined, they were merged and continuum points from the continuum grid were added outside line groups.

\subsection{Background opacity}\label{backgr_opac}
Opacity and emissivity of transitions were computed for each atom using Eqs. \ref{eq:opacity}, \ref{eq:RTCl}, and \ref{eq:RTCc}. The photoionisation cross-section of hydrogen-like ions was computed using Kramer's formula with correction from quantum mechanics (HM14, Eq. 7.92).
For non-hydrogen-like ions, the photoionisation cross-section was directly read from the atomic file and interpolated on the wavelength grid of MCFOST. We took the photoionisation data from TOPbase\footnotemark \citep{topbase}.
\footnotetext{\url{http://cdsweb.u-strasbg.fr/topbase/topbase.html}}

We also included continuum transitions from other sources which add up in $\chi_c$. These are Thomson scattering on free electrons, hydrogen free-free (HM14, Eq. 7.100), and \textsc{H$^{{\rm-}}$} free-free and bound-free transitions from \cite{John1988}.

\subsection{Non-LTE loop}\label{nonLTE_loop}
\begin{figure*}
\centering
  \includegraphics[width=0.8\textwidth]{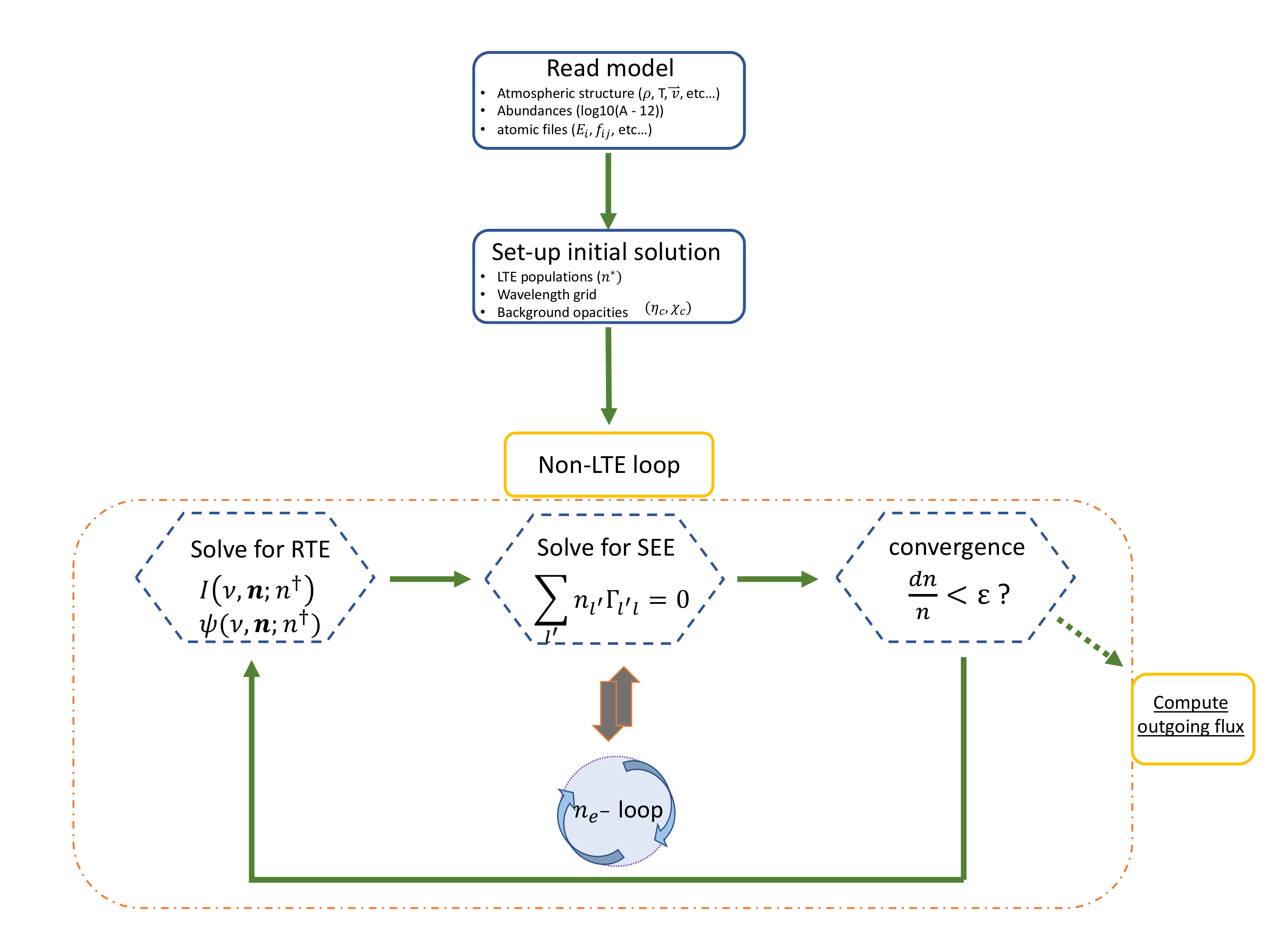}

  \caption{Flow chart of MCFOST-art. See \S \ref{nonLTE_loop} for more details. The green arrows indicate the main iterative loop. The double grey arrow indicates that the electron density loop is iterated within the main iterative loop and with the current estimates of the non-LTE populations.}
    \label{fig:flowchart}
\end{figure*}

The main module of our code deals with solving Eq. \ref{eq:SEE2} simultaneously with Eq. \ref{eq:RTE} via the non-LTE loop. Figure \ref{fig:flowchart} shows the main steps leading to the solution of the non-LTE problem with MCFOST-art. We stress that even if Eq. \ref{eq:SEE2} is solved locally (i.e. within each cell) it is coupled with other cells through Eq. \ref{eq:RTE}.

Firstly, the code reads abundances for the atomic species under consideration, the corresponding atomic data  (e.g. energy levels $E_i$ and the oscillator strength $f_{ij}$), and an atmospheric model (temperature, densities, and velocity fields). The electron density is evaluated at LTE if not present in the model (for more detail on electron density loop, see sect. \ref{sect:electron}). Secondly, MCFOST-art initialises the level populations of each atom as described above.

Finally, we evaluate continuous background opacities. In models in which electron scattering is an important source of opacity, the continuum mean intensity and the corresponding Thomson scattering emissivity are computed with the ALI method (for more details, see chapter 13 of HM14).
Once background opacities have been evaluated, the non-LTE loop starts. It computes a self-consistent solution for level populations and radiative transfer equations. For each cell, the rate matrix $\Gamma_{\lprime \l}$ is built for a set of rays, for all wavelengths simultaneously. 
Once the rate matrix is known, a new value of the level populations is determined.
Electron densities can be evaluated with the new values of the non-LTE populations. To avoid convergence issues, the evaluation of the electron density can be done every N iterations of the non-LTE loop. 

The new populations are then compared to the previous populations using the following criterion of convergence:
\begin{equation}\label{eq:convergence}
    \dfrac{\delta n}{n} = \| 1 - \dfrac{n^{\dagger}}{n} \| < \epsilon \text{,}
\end{equation}
where $\epsilon$ is a user defined convergence threshold. The threshold $\epsilon$ is usually chosen between 10$^{-3}$ - 10$^{-4}$.
If the maximum relative changes for each cell and each atom is below the convergence threshold the non-LTE loop stops. Otherwise, a new iteration starts with the previously computed populations.
Excitation and ionisation temperatures for each transition of each atom are also checked for convergence. In general, transition temperatures converge faster than level populations.

\subsection{Electron density}\label{sect:electron}

Electron densities for a gas mixture of hydrogen and metals cannot be determined analytically. 
The populations of atomic levels rely on the knowledge of the electron densities (for instance, through the Saha-Boltzmann equation at LTE). In turn, the electron density is a function of the population of atomic levels through their ionisation fraction. Therefore, electron densities have to be determined iteratively. 
Starting from an initial guess for the electron density, the populations of atomic levels are derived and the new ionisation fractions $f_{is}$ for level $i$ in the ionisation stage $s$ for each electron donors are evaluated. Then, from the new ionisation fractions, electron densities are computed.
This iterative procedure is repeated until the relative change in electron densities drops below $10^{-6}$.

In our code we use a linearisation procedure similar to Eq. 17.88 of HM14. The initial guess for the electron density is given by the ionisation of hydrogen plus one metal (M) $n_e^0 = n_e(\textsc{H}) + n_e(\textsc{M})$. Hydrogen ionisation yields
\begin{equation}\label{eq:neH}
    n_e(\textsc{H}) = (\sqrt{n_H \,\phi_H + 1}-1)/\phi_H \text{,}
\end{equation}
and the metal ionisation
\begin{equation}\label{eq:neM}
    n_e(\textsc{M}) = (\sqrt{\alpha_M\, n_H\,\phi_M +\dfrac{1}{4}\,(1+\alpha_M)^2}
     - \dfrac{1}{2}\,(1+\alpha_M) ) / \phi_M \text{,}
\end{equation}
where $n_H$ is the total number of hydrogen atoms, $\phi_x$ the Saha-Boltzmann factor for the ion x, and $\alpha_M$ is the abundance of the metal relative to hydrogen. 

\subsection{Collisional rates}\label{sect:Collision}

The evaluation of the collisional rates appearing in the SEE (Eq. \ref{eq:SEE}) is cumbersome. It generally requires collision calculations with quantum mechanics for which look-up tables of collisional rates are available now \citep{Barklem2016a}.
However, analytical and semi-analytical recipes exist for faster evaluation of the collision matrix $C_{ji}$ (which is proportional to the collisional rates within a factor).
We consider collisional excitation and ionisation by electrons for hydrogen atoms using \cite{Johnson1972}. For collisional excitation of metals (and helium) by electrons we use van Regemorter's fomula \citep{Regemorter1962} and the impact parameter method \citep{Seaton1962}. For ions, we follow Eq. 34 of \cite{Landolt-Bornstein-v2} and, for neutrals, we follow \cite{Seaton1962}. For collisional ionisation by electrons, we use Eq. 9.60 of HM14 and  for collisions with hydrogen atoms, we use Drawin's formula (Eq. 9.62 of HM14).

\subsection{Quadratures}\label{angle_quad}
One of the cornerstones in solving the SEE is the evaluation of the integrals appearing in Eq. \ref{eq:RR}. In the following, we describe how we performed wavelength and angular integrations.

Angular quadrature for molecular lines transfer in MCFOST is done with the Monte Carlo method. For atomic lines, we used a set of fixed directions and weights uniformly distributed on the unit sphere such that integration over the sphere gives $4 \pi$. We used the directions and weights of type A quadratures from B. Carlson (whose calculations are described in \cite{Bruls99}) and the directions and weights from \cite{Stepan2020} for unpolarised radiation. These two types of quadrature points give similar results. By default we used type A quadrature of Carlson with 80 points sampling $4 \pi$ steradians \citep{Ibgui2013}.

\cite{DeCeuster2020} used a similar angular quadrature method but with the quadrature points and weights determined by the HEALPix algorithm \citep{HEALPix}. For each cell of a model, starting from the centre of the cell (hereafter one point quadrature), the transfer equation is solved for all directions, while the radiative rates and the mean intensity are accumulated. Integrating the transfer equation from the centre of each cell may underestimate angular means and introduce some inaccuracy in the level populations because it only gives the mean intensity at that position. However, we do not know beforehand whether this inaccuracy is significant or not. Thus, we implemented two options to deal with this issue: first, several starting points for the angular quadrature scheme can be selected for each cell; and, second, Monte Carlo iterations can be done to properly sample each cell, in addition to the one point angular quadrature described above.
In \S \ref{accurate_pops}, we discuss these two options.

We replace the double integral appearing in Eq. \ref{eq:Jbar} by a sum over wavelength points and directions as follows:
\begin{equation}
    \displaystyle\int \dfrac{d\Omega}{4\pi} \displaystyle\int d\nu\, I(\nu, \mathbf{n})\,\phi (\nu, \mathbf{n}) \equiv \displaystyle\sum\limits_{{\rm ray}} \omega_{{\rm ray}} \displaystyle\sum\limits_{\lambda} 
    \omega_{\lambda} I(\lambda, {\rm ray})\,\phi (\lambda, {\rm ray} )\text{,}
\end{equation}
where $\omega_{{\rm ray}}$ is the weight of the angular quadrature for a given direction (or ray) and $\omega_{\lambda}$ the weight of a trapezoidal wavelength quadrature.
In the case of a line transition, $\omega_{\lambda}$ includes the normalisation of the line profile.

\section{Benchmarking the code}\label{benchmarks}

The MCFOST code is intrinsically 3D, but in this paper we focus on 1D models that test the core of the method and allow us to validate our implementation with well-established 1D codes. While this is not the main application intended for our code, it also illustrates that MCFOST can now be used to generate a variety of stellar spectra. We will devote a forthcoming paper to the applications of our code to 2D and 3D geometries, with a particular emphasis on stellar magnetospheres.
We used the codes TURBOspectrum \citep{turbospeccode} and RH \citep{Uitenbroek2001} as references for testing.

The code MCFOST does not include a dedicated grid for 1D models, and we used the spherical grid instead. However, the spherical grid of MCFOST is log-scaled in the radial direction and cannot efficiently map the non-uniform grid from the 1D models. Therefore, for our benchmarks, we directly read the grid from the input models as RH and TURBOspectrum do. In that case, a cell is defined as the distance between two consecutive grid points. Because Eq. \ref{eq:RTE} is solved in MCFOST by summing up the intensity of each cell (i.e. spatial grid unit) weighted by the optical depth (see Eq. \ref{eq:RTEmcfost}), the solution is more dependent on the grid resolution (i.e. on the size of cells). On the contrary, RH and TURBOspectrum formal solvers typically interpolate the source function in the initial grid, resulting in a solution that is highly accurate. To mitigate the impact of a lack of resolution on the accuracy of the solution, we decreased the size of the cells by adding more points between each point of the initial models.

The TURBOspectrum code is used to model spectra from cool evolved stars \citep{turbospectrum1,turbospectrum2} in synergy with the model atmosphere code MARCS \citep{MARCS2008}. This code works only at LTE but includes a wide library of molecules, the occupation probabilities formalism, and a proper treatment of hydrogen line broadening. To match the capabilities of our code, we slightly modified TURBOspectrum and removed molecular opacities, although, at the coolest temperature, neutral hydrogen density is dependent on the formation of molecules, such as \textsc{H}$_2$.

The radiative transfer code RH is mainly used to model line formation in non-LTE conditions using the MALI method. Recent versions of RH have been used to model spectral lines taking into account the Zeeman effect and partial frequency redistribution in 3D cubes of the solar photosphere \citep{rh_stic}. Recently, \cite{Criscuoli2020} benchmarked the RH code and showed its ability to model the solar irradiance with a good agreement with observations.
The RH and MCFOST codes use similar treatment of background opacities and both employ the MALI method to solve for level populations.
We compared our treatment of background opacities and level dissolution at LTE with TURBOspectrum. We used RH for benchmarking the MALI method on the non-LTE formation of hydrogen lines.
We used the (1D) spherically symmetric grid of TURBOspectrum and RH for all tests.

\begin{table*}
\centering
\begin{tabular}{cccccc}
\hline
\hline
Model ID & ${\rm T_{eff}}$ (K) & ${\rm \log g \, (cm\,s^{-2})}$ & radius (${\rm R_{\odot}}$) & ${\rm \xi \, (km\,s^{-1})}$ & LTE  \\ 
\hline
\\
sun & 5777 & 4.44 & 1.00 & 1 & yes \\
s8000 & 8000 & 2.50 & 9.34 & 1 & yes \\
s3500 & 3500 & 3.50 & 2.95 & 2 & yes  \\
\\
C & 5777 & 4.44 & 1.00 & variable & no  \\
\\
\hline
\end{tabular}
\caption{Stellar atmosphere models used for the benchmarking of our code. Models are taken from the MARCS database and correspond to the photosphere, except the model C taken from \cite{FALC}. The first column indicates the designation of the model; the second, third, and fourth are the effective temperature, log surface gravity, and radius of the star, respectively. The fifth column gives the constant microturbulence, except for the FAL-C model, which has a depth varying microturbulence. The last column indicates if the model is computed assuming LTE.}
\label{tab:models_para}
\end{table*}

\subsection{Stellar photospheres}

To test our opacity modules and background opacity calculations, we used LTE models of stellar photospheres from the MARCS database\footnotemark \citep{MARCS2008}.
\footnotetext{\url{https://marcs.astro.uu.se/}} The summary of each model is given in Table \ref{tab:models_para}.
The first model, $sun$, is the standard solar photosphere model from the MARCS code. The last two models, $s8000$ and $s3500$, are representative of the photospheres of hot and cool giants, respectively. 
All models have standard chemical composition and are computed assuming spherical symmetry, except for the solar model, which was computed using the plane-parallel approximation.

\subsubsection{Continuum radiation}\label{star_continuum}

\begin{figure*}[hbtp!]
\begin{subfigure}{0.5\textwidth}
  \centering
  \includegraphics[width=1\textwidth]{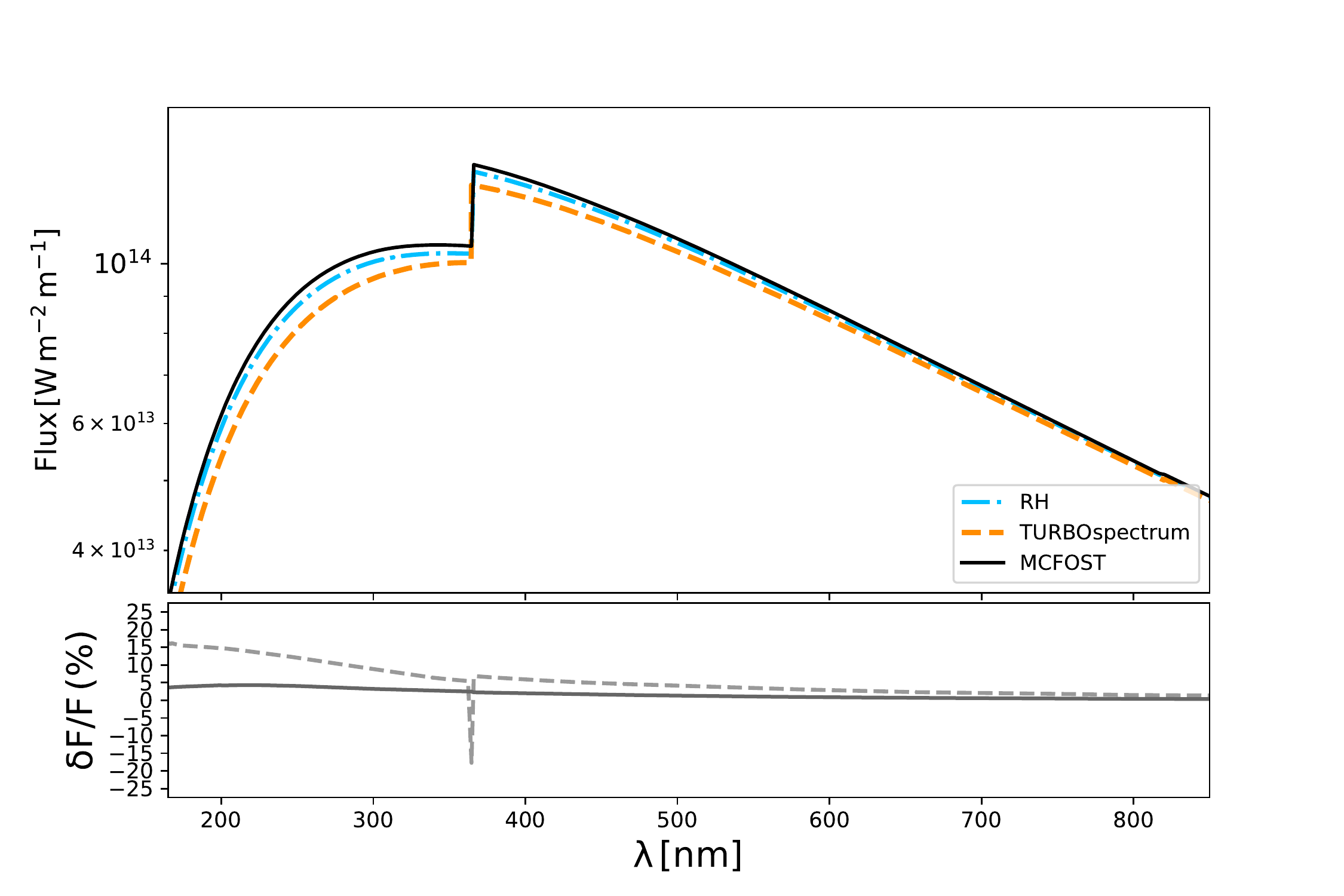}
 \caption{Solar photosphere without level dissolution.}
\end{subfigure}
\begin{subfigure}{0.5\textwidth}
  \centering
  \includegraphics[width=1\textwidth]{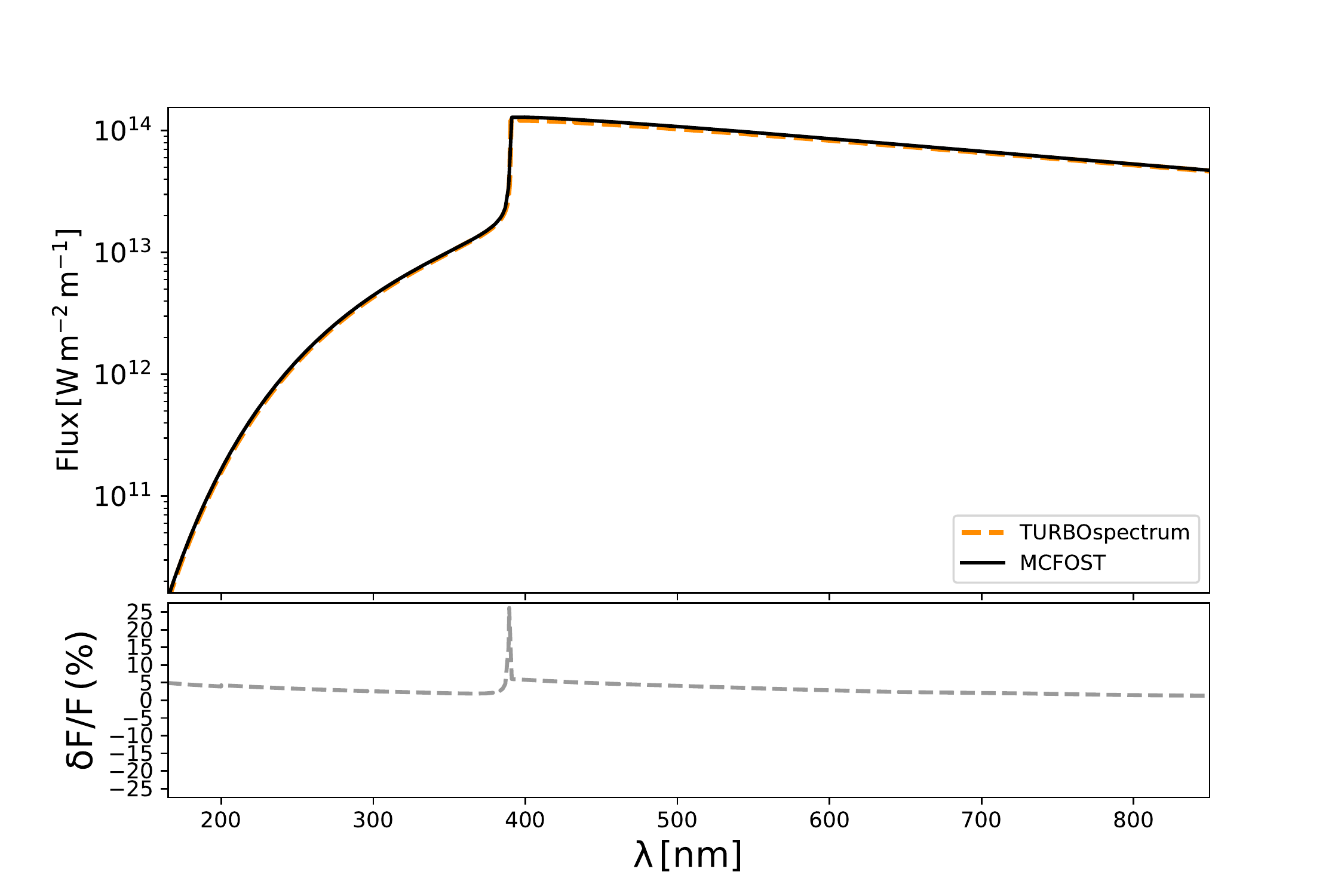}
 \caption{Solar photosphere with level dissolution.}
\end{subfigure}
\newline
\begin{subfigure}{.5\textwidth}
  \centering
  \includegraphics[width=1\textwidth]{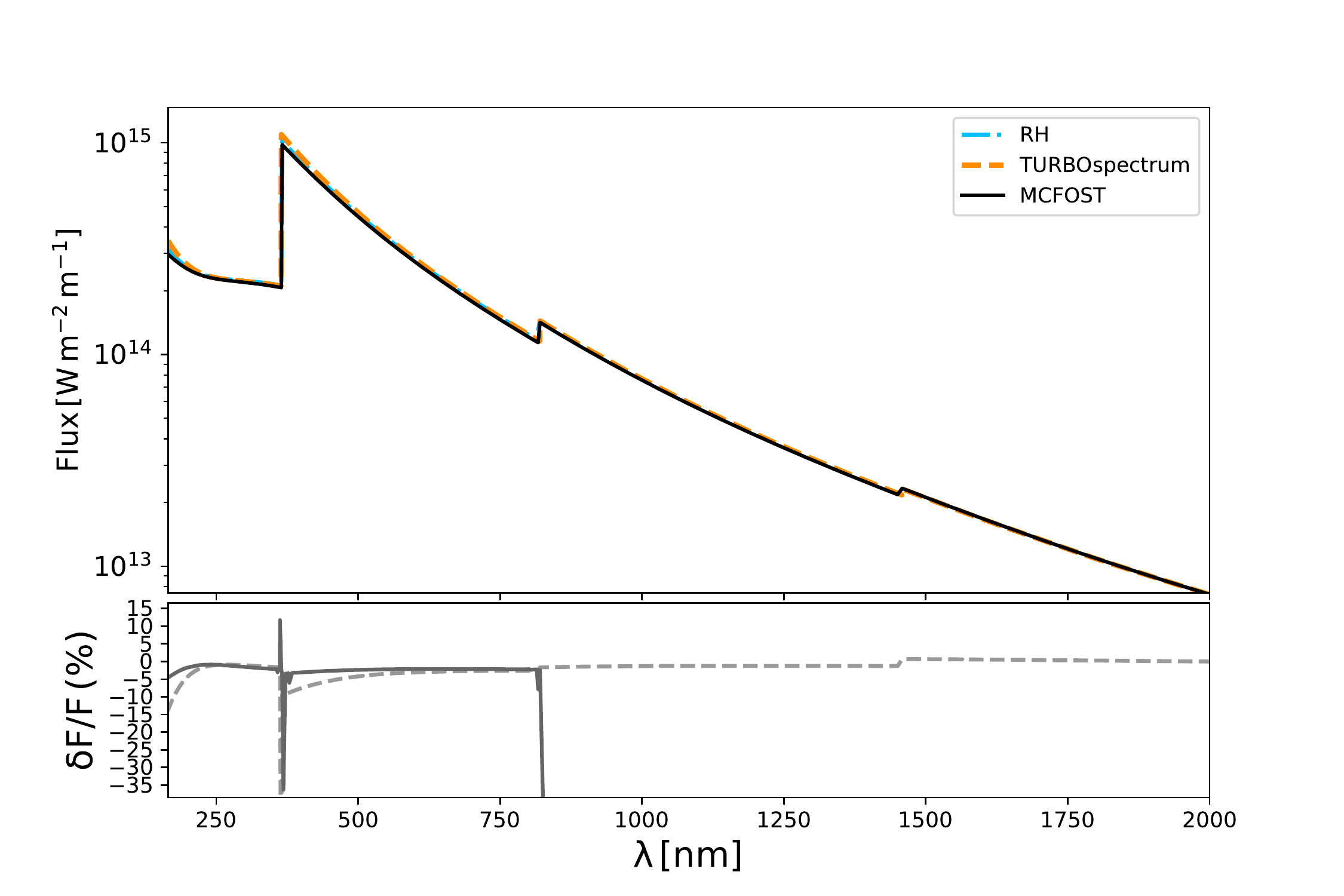}
 \caption{$s8000$ model without level dissolution.}
\end{subfigure}
\begin{subfigure}{.5\textwidth}
  \centering
  \includegraphics[width=1\textwidth]{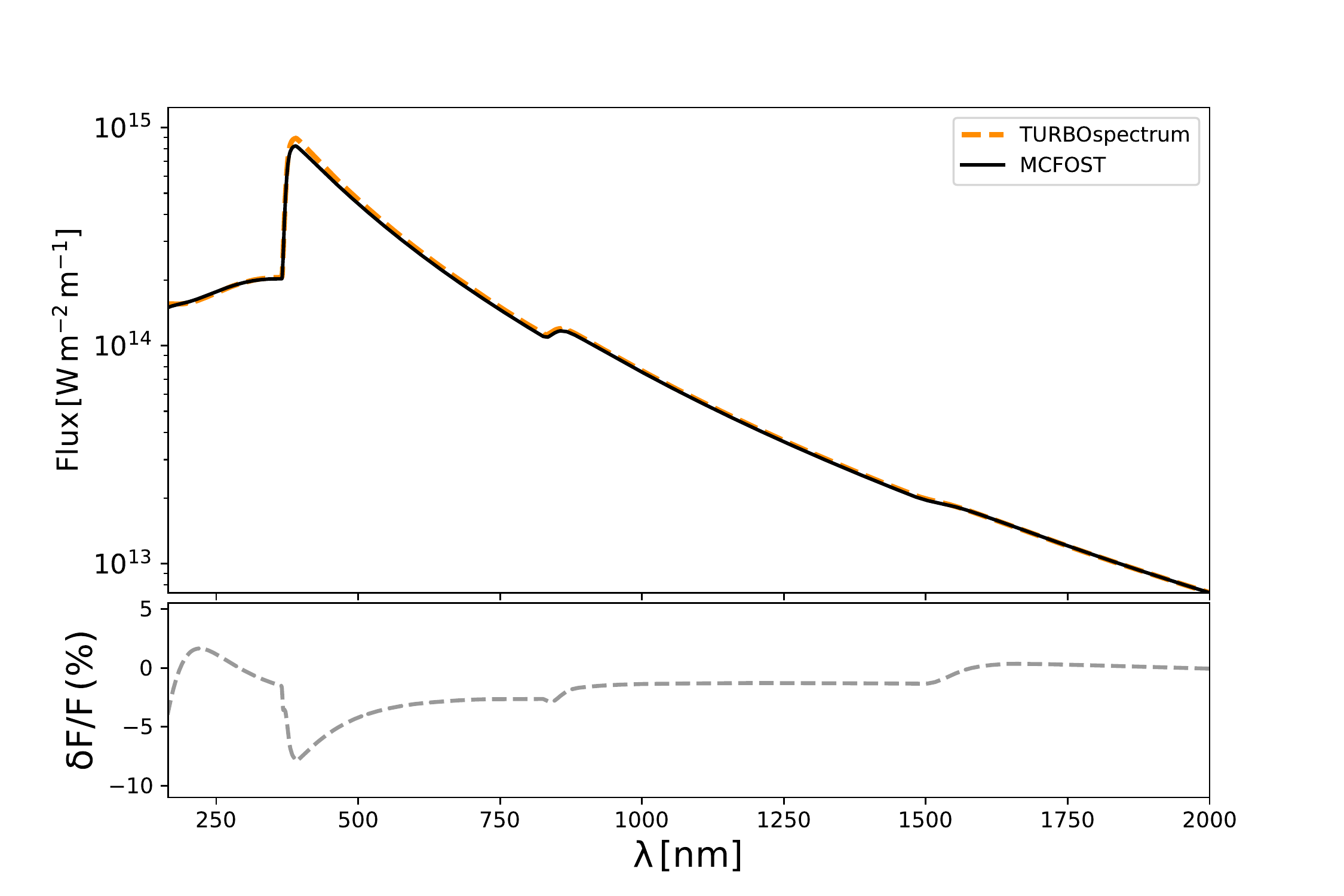}
 \caption{$s8000$ model with level dissolution.}
\end{subfigure}
\newline
\begin{subfigure}{.5\textwidth}
  \centering
  \includegraphics[width=1\textwidth]{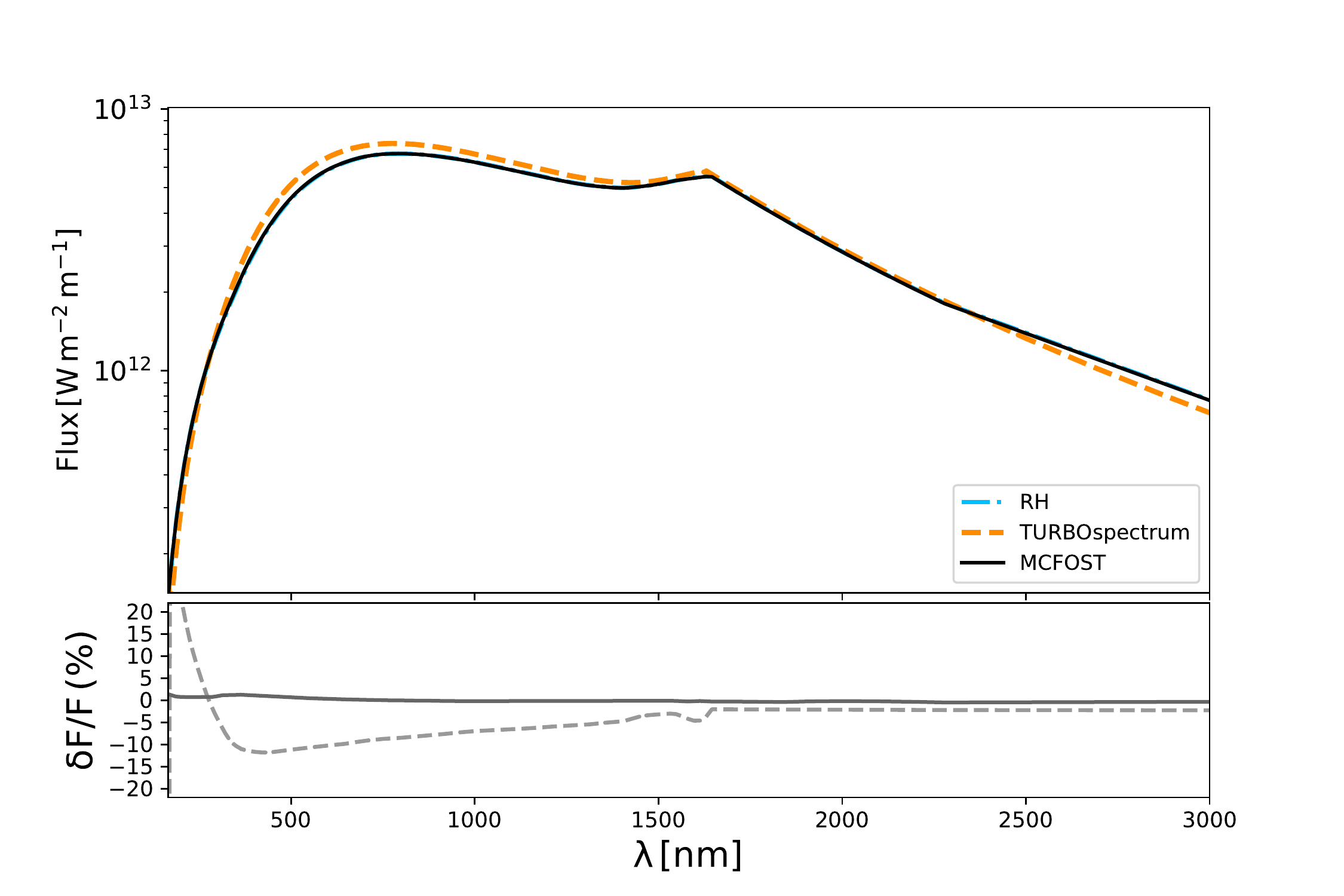}
 \caption{$s3500$ model without level dissolution.}
\end{subfigure}
\begin{subfigure}{.5\textwidth}
  \centering
  \includegraphics[width=1\textwidth]{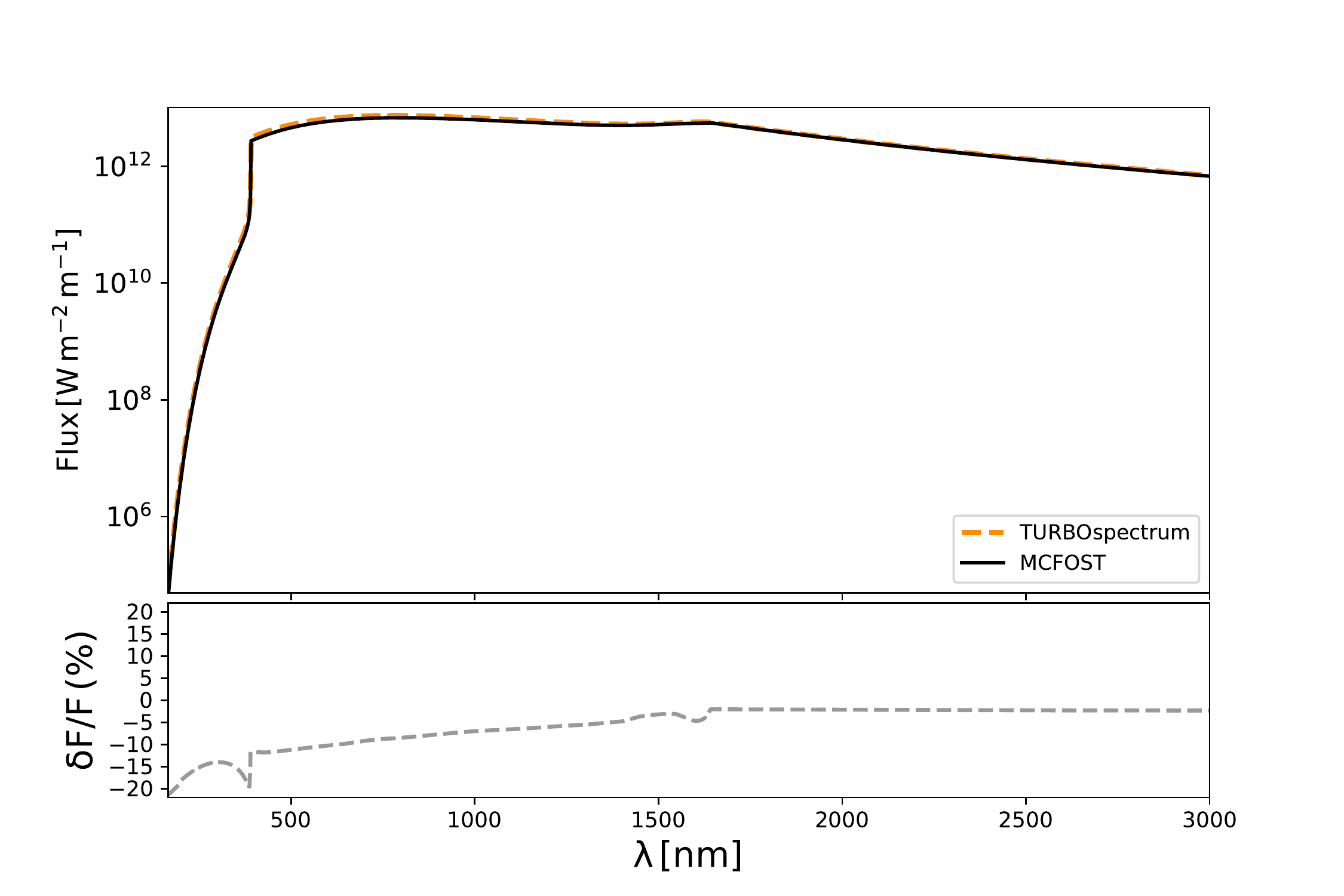}
 \caption{$s3500$ model with level dissolution.}
\end{subfigure}

  \caption{Continuum flux as a function of wavelengths for the photospheric models without (left panels) and  with (right panels) level dissolution. 
 In each panel, the upper window shows the flux from RH, TURBOspectrum, and MCFOST in cyan (thin line), orange (dashed), and black (thick line), respectively. The lower window shows the difference between RH and MCFOST  (thick dark grey lines), and between TURBOspectrum and MCFOST (dashed light grey lines).
  As RH does not include level dissolution, it is not shown in the right figures. In figure 2c, the continuum flux computed by RH stops at about 800 nm, leading to incorrect evaluation of the error at this edge point.
 }
    \label{fig:cont_fluxes}
\end{figure*}

We compared the stellar continuum computed by the three codes for the stellar photosphere models (see Table \ref{tab:models_para}).
Figure \ref{fig:cont_fluxes} shows the continuum flux and the discrepancy function $\delta F/F$ for each model.

The agreement between RH and MCFOST is excellent, being at or below 1\% at most wavelengths. For models $sun$ (Fig. \ref{fig:cont_fluxes}a) and $s3500$ (Fig. \ref{fig:cont_fluxes}e), the largest discrepancies are due to interpolation errors at the shortest wavelengths. For the model $s8000$ (Fig. \ref{fig:cont_fluxes}c) the agreement between RH and MCFOST is also at the percent level, except at the Balmer jump (about 350 nm), where differences are larger (again owing to the steep slope of the emission), but remain below 30\%.
Discrepancies between MCFOST and TURBOspectrum are of the same order as those between MCFOST and RH. 
The largest discrepancy for the model $s3500$, of the order of 10\%, is due to the \textsc{H}$^-$ continuum. In TURBOspectrum, the \textsc{H}$^-$ density is self-consistently determined by solving chemical equilibrium equations including hydrogen in its atomic and molecular forms, leading to a slightly different value than in RH and MCFOST.
For the model $s8000$, the largest discrepancy occurs at the Balmer jump and is of the order of 40\%. As the largest discrepancy with RH occurs at the same location we might not exclude numerical resolution problems. The agreement between RH and TURBOspectrum is of the same order as MCFOST and TURBOspectrum. Furthermore, we note that for this model, continuum scattering becomes non-negligible and is automatically included in TURBOspectrum\footnotemark, but not in MCFOST, thereby impacting the shape of the continuum.

\footnotetext{Continuum scattering cannot be removed  easily in TURBOspectrum to benchmark further the codes, but it is negligible for all models but the hotter model.}

The agreement between MCFOST and TURBOspectrum remains excellent when level dissolution is taken into account (Fig. \ref{fig:cont_fluxes} right panels).
The impact of level dissolution is clearly visible at the Balmer jump where it smoothes the abrupt discontinuity towards redder wavelengths. The shallower spectrum results in an improved agreement between TURBOspectram and MCFOST for model $s8000$ around the Balmer jump.

\subsubsection{Velocity fields}

Although the impact of velocity fields on line formation will be addressed in a forthcoming paper, in this section we show an example of a solar photosphere with a radial velocity $V_r$ given by

\begin{equation}\label{eq:wind}
V_{r} = v_{0} + (V_{\inf} - v_{0} ) \, (1 - r_{0}/r)^{\beta} \text{,}
\end{equation}
where $v_{0}$ is the velocity at the inner point (bottom of the photosphere), $V_{\inf}$ a velocity such that ${\rm max(V_r)  = 500\, km\,s^{-1}}$, and $\beta$ the exponent of the velocity law. In our calculations, ${\rm v_{0} = 9\,km\,s^{-1}}$ and $\beta = 0.5$.

We compared the result of our code with RH since TURBOspectrum does not deal with macroscopic velocity fields. For comparison, we implemented a solver for the velocity field in RH because the spherically symmetric version does not handle velocity fields. However, unlike MCFOST, the velocity fields between two points are not interpolated.

\begin{figure}[H] 

  \includegraphics[width=1\columnwidth]{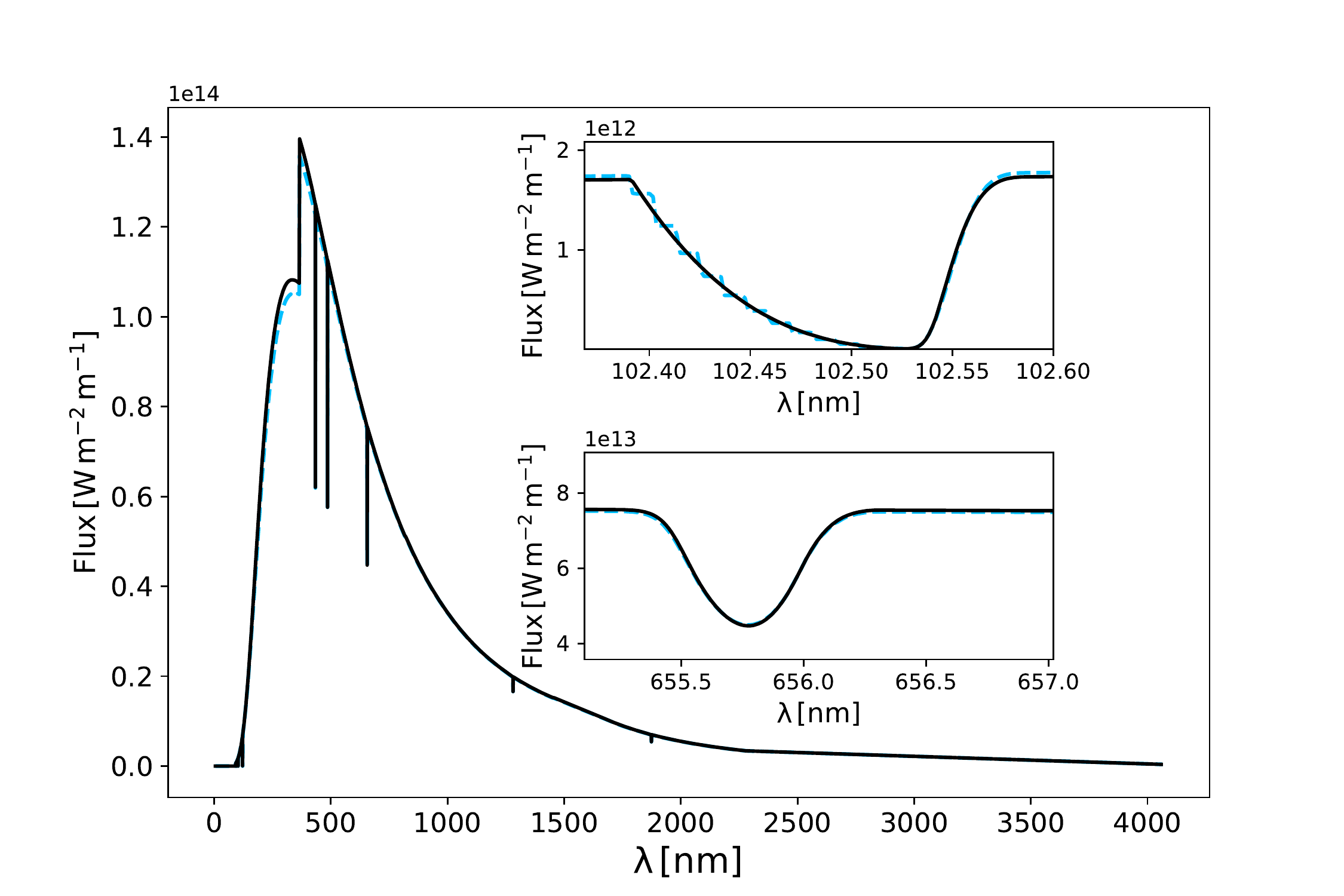}

  \caption{Flux for a solar photospheric model with velocity gradient given by Eq. \ref{eq:wind}. The flux from MCFOST is shown in black and from RH in cyan. A zoom-in around Ly$\beta$ and \textsc{H}$\alpha$ is also shown.}
    \label{fig:flux_sun_vel}
\end{figure}

Figure \ref{fig:flux_sun_vel} shows the difference between MCFOST and RH for the solar photosphere when a velocity field is included. While the continuum radiation is similar to that of Fig. \ref{fig:cont_fluxes}a, lines are impacted significantly by the large velocity gradient, as expected. The differences between the two codes are negligible, although the lack of resolution in RH is noticeable.

\begin{figure}[H]

  \includegraphics[width=1\columnwidth]{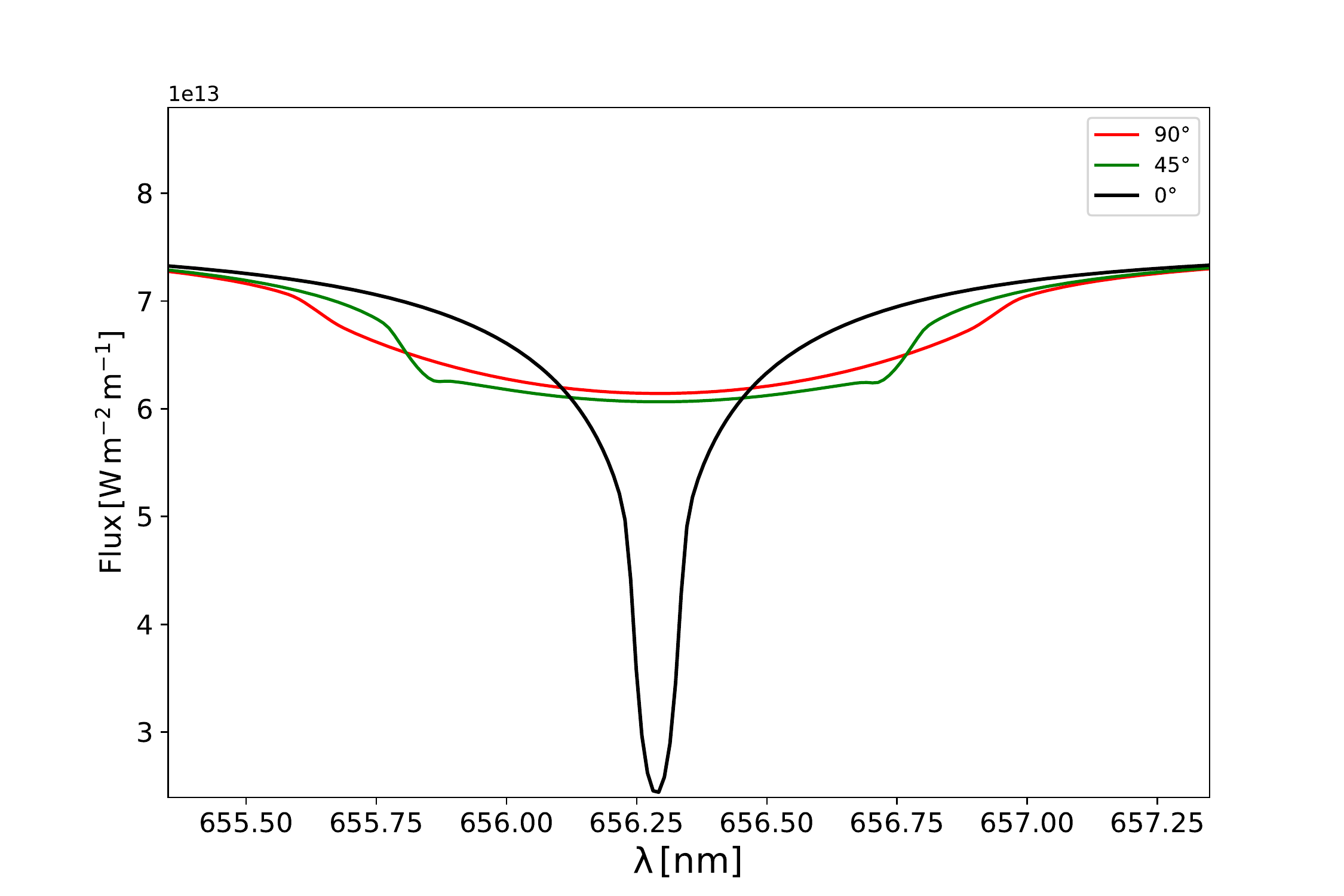}

  \caption{\textsc{H}$\alpha$ flux variation with inclination for a solar photospheric model with rotation. The flux at inclinations of $0^{\circ}$, $45^{\circ}$, and $90^{\circ}$ are shown in black, green, and red, respectively.
  An inclination of $0^{\circ}$ means that the star is seen pole on.}
    \label{fig:flux_sun_rot}
\end{figure}

We also tested the treatment of the same model with no radial velocity but a constant rotational velocity of 300 ${\rm km\,s^{-1}}$ instead. Figure \ref{fig:flux_sun_rot} shows the \textsc{H}$\alpha$ line flux as a function of inclination. As expected, the impact of rotation on the line shape increases with inclination, from no effect when the star is seen pole-on to maximum broadening when the star is equator-on.

\subsection{Solar atmosphere model}\label{solar_atm}

Although a LTE model of the solar photosphere is adequate to model most solar lines, it fails to reproduce lines formed beyond the photosphere. The upper solar atmosphere is a high temperature and low-density region above the photosphere where non-LTE effects dominate the line formation. 
In the following we focus on the Lyman $\alpha$ (Ly$\alpha$) and \textsc{H}$\alpha$ lines as they form in the upper atmosphere \citep[e.g. Figure 1 of][]{Vernazza81} and are thus impacted by non-LTE effects.

We used a 6-level hydrogen atom, with five bound levels and one continuum level, which is the ground state of \textsc{H} II.
Energy levels and transition frequencies are from \cite{Johnson1972}. Bound-free cross-sections are computed using Kramer's formula and collisional rates are evaluated following \cite{Johnson1972}. 
For the Lyman $\alpha$ and \textsc{H}$\alpha$ lines, the absorption profiles are given by a Voigt function. For the other lines, we used a Gaussian profile.
We used model C of \cite{FALC} as a standard model of the solar atmosphere including the chromosphere and transition region.

\begin{figure*}[hbtp!]
  \centering

  \includegraphics[width=\textwidth]{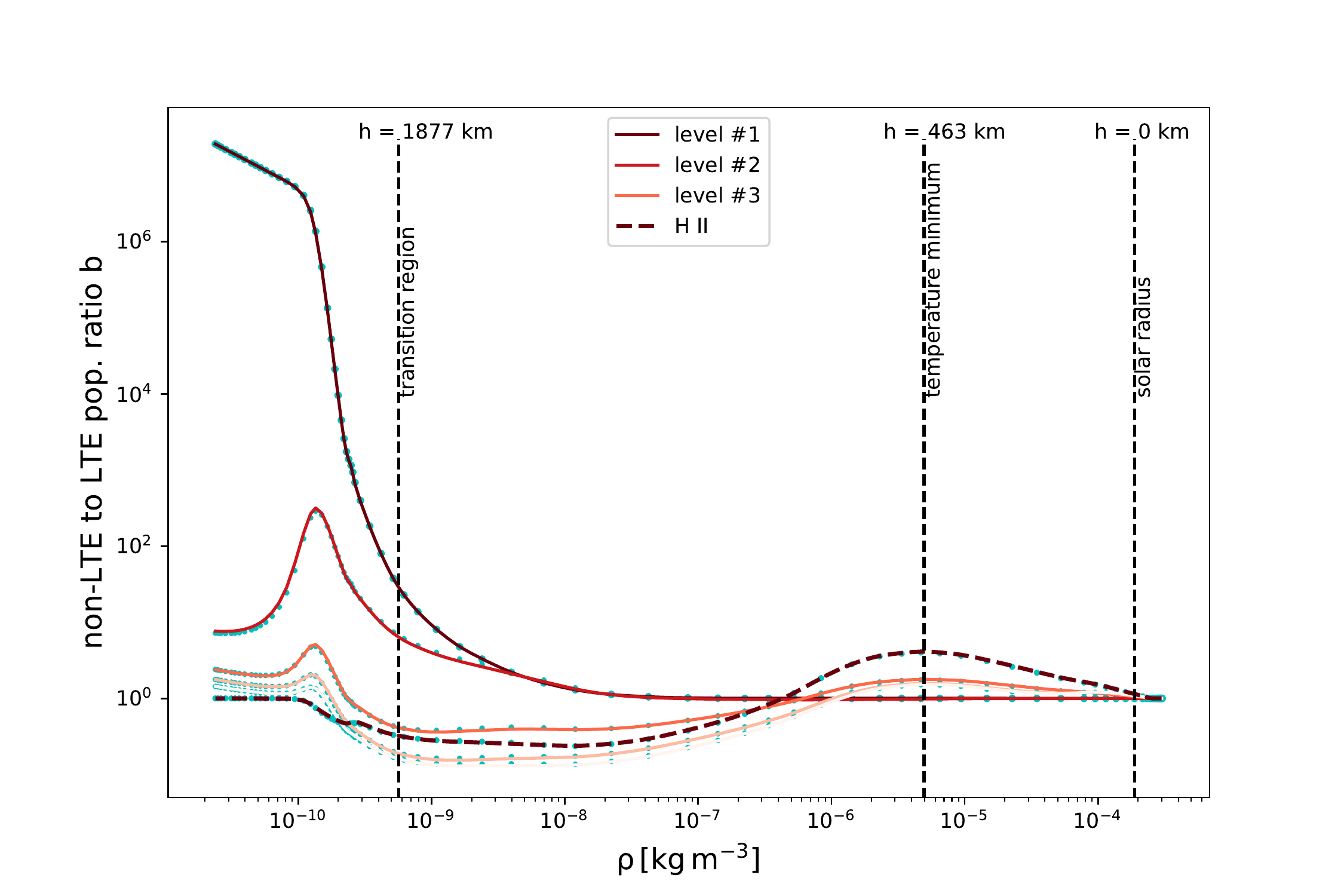}

  \caption{Ratio of non-LTE to LTE populations (departure coefficient b) of a 6-level hydrogen atom as a function of density ($\rho$). RH populations are indicated with cyan dots. MCFOST populations are shown with full colour lines, where darker (lighter) red shades mean lower (higher) energy levels. For clarity, the first three atomic levels and the continuum level (ground state of \textsc{H}II) are labelled explicitly in the insert.
  The different layers of the solar atmosphere are shown with vertical dashed lines. Typical heights above the solar radius are also indicated.}
    \label{fig:pops_falc_b}
\end{figure*}

\begin{figure*}[hbtp!]
\begin{subfigure}{0.5\textwidth}
  \centering

  \includegraphics[width=1.1\textwidth]{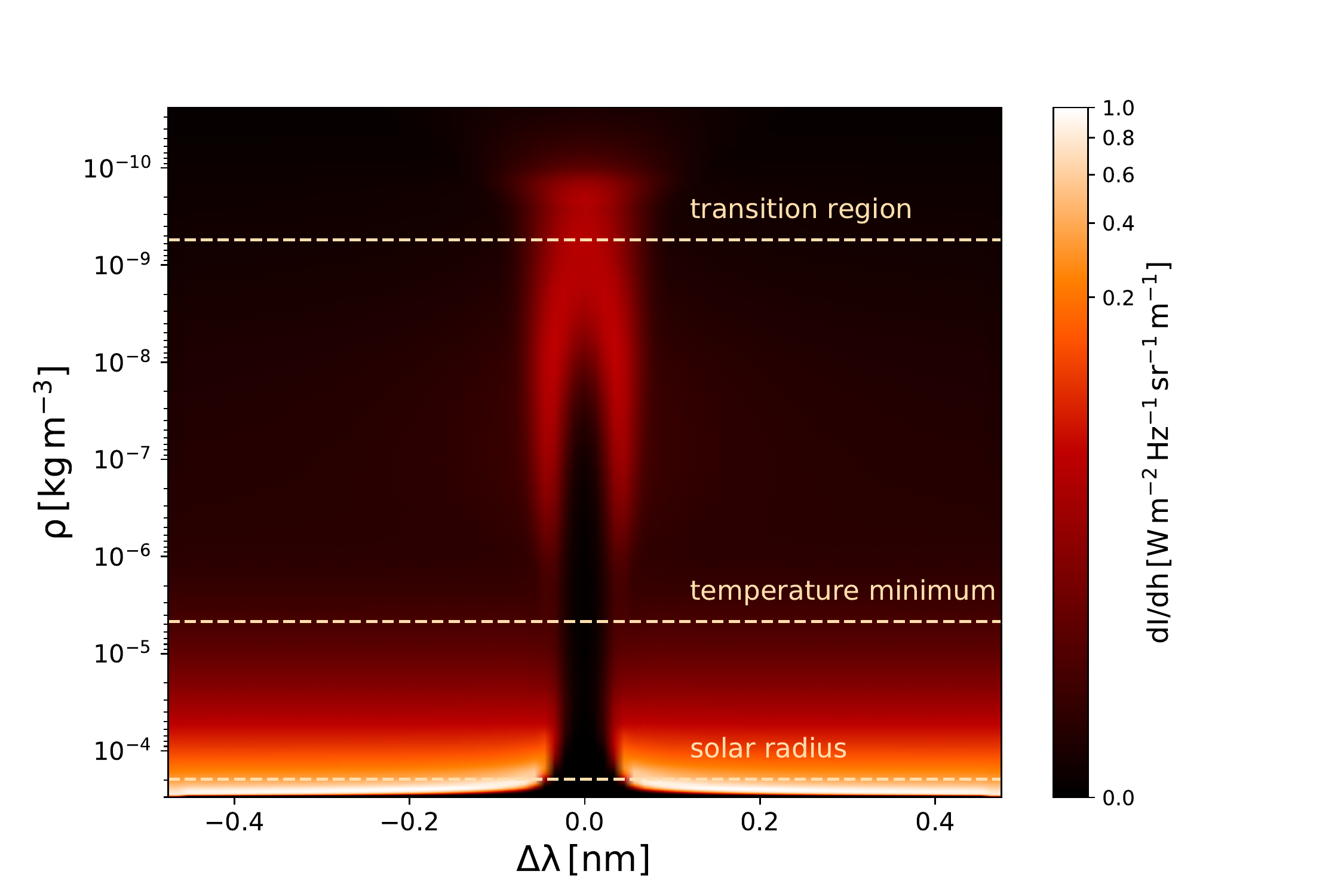}
  
 \caption{\textsc{H}$\alpha$ line.}
\end{subfigure}
\begin{subfigure}{0.5\textwidth}
  \centering

   \includegraphics[width=1.1\textwidth]{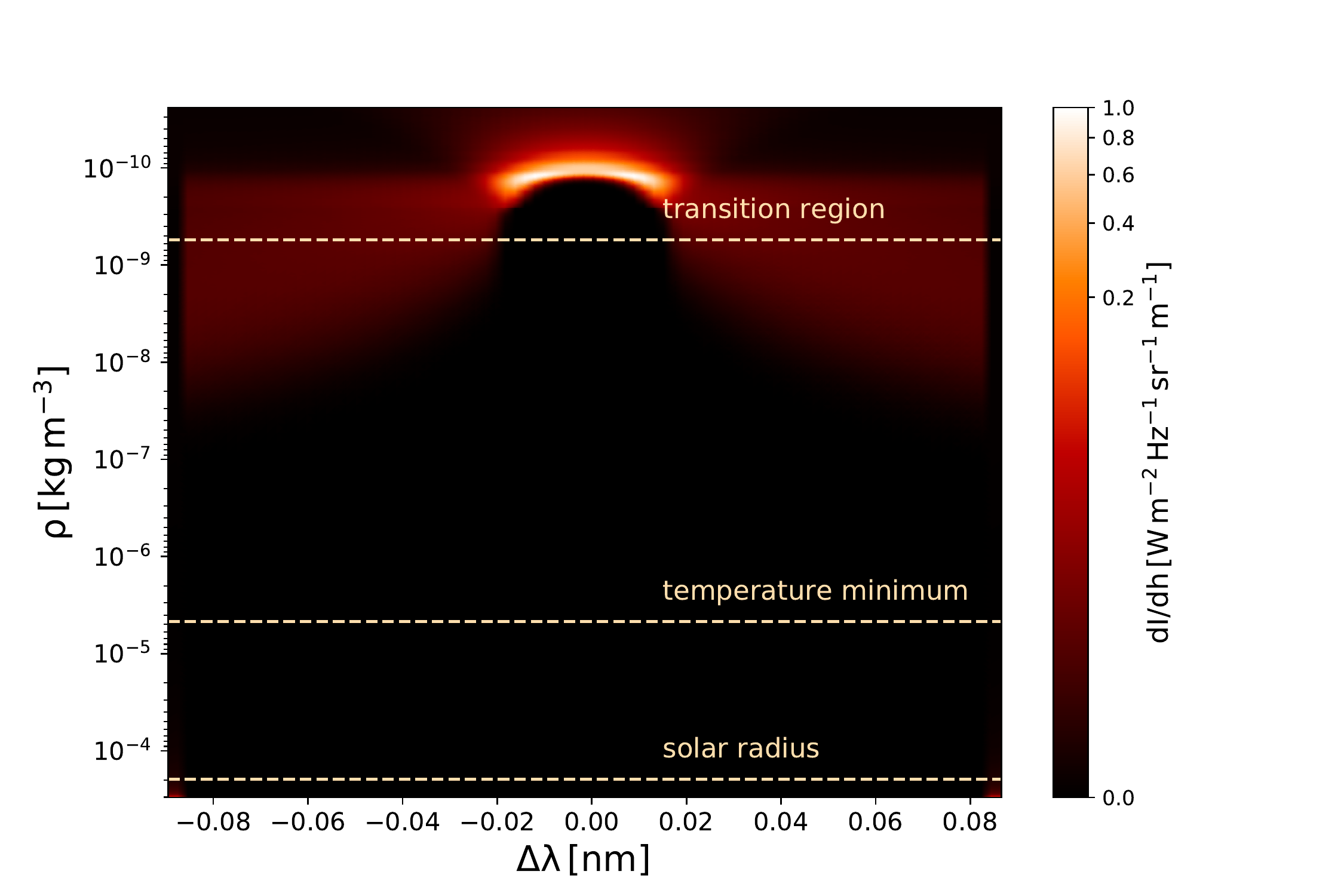}
  
 \caption{Ly$\alpha$ line.}
\end{subfigure}

  \caption{Line contribution function, $\dfrac{dI}{dh}$ (response function to height variation), at disc centre.  The contribution function is drawn as a function of the density in the atmosphere and distance from line centre (in nm). Lighter areas show where the contribution function peaks. 
  We indicate the different layers of the solar atmosphere in each panel.}
    \label{fig:contrib_functions}
\end{figure*}

Overall, the agreement between RH and MCFOST is excellent.
Figure \ref{fig:pops_falc_b} shows the ratio $b$ between non-LTE and LTE populations in model C. In the photosphere (below the temperature minimum), the populations are mostly at LTE with a $b$ ratio of a few. In the deep photosphere, $b$ is close to 1. From the temperature minimum to the transition region (the chromosphere), the populations start to depart from their LTE values. In the lower corona, the populations are very far from LTE, with a $b$ factor greater than $10^{6}$ for the ground state of \textsc{H}I (level 1 in Fig. \ref{fig:pops_falc_b}). The formation regions of the two lines are consistent with the earlier work of \cite{Vernazza81}. Contribution functions (i.e. regions of formation) of the Ly$\alpha$ and \textsc{H}$\alpha$ lines are shown in Figure \ref{fig:contrib_functions}. 

\begin{figure}[H]

\begin{subfigure}{1\columnwidth}
  \centering
   \includegraphics[width=1\columnwidth]{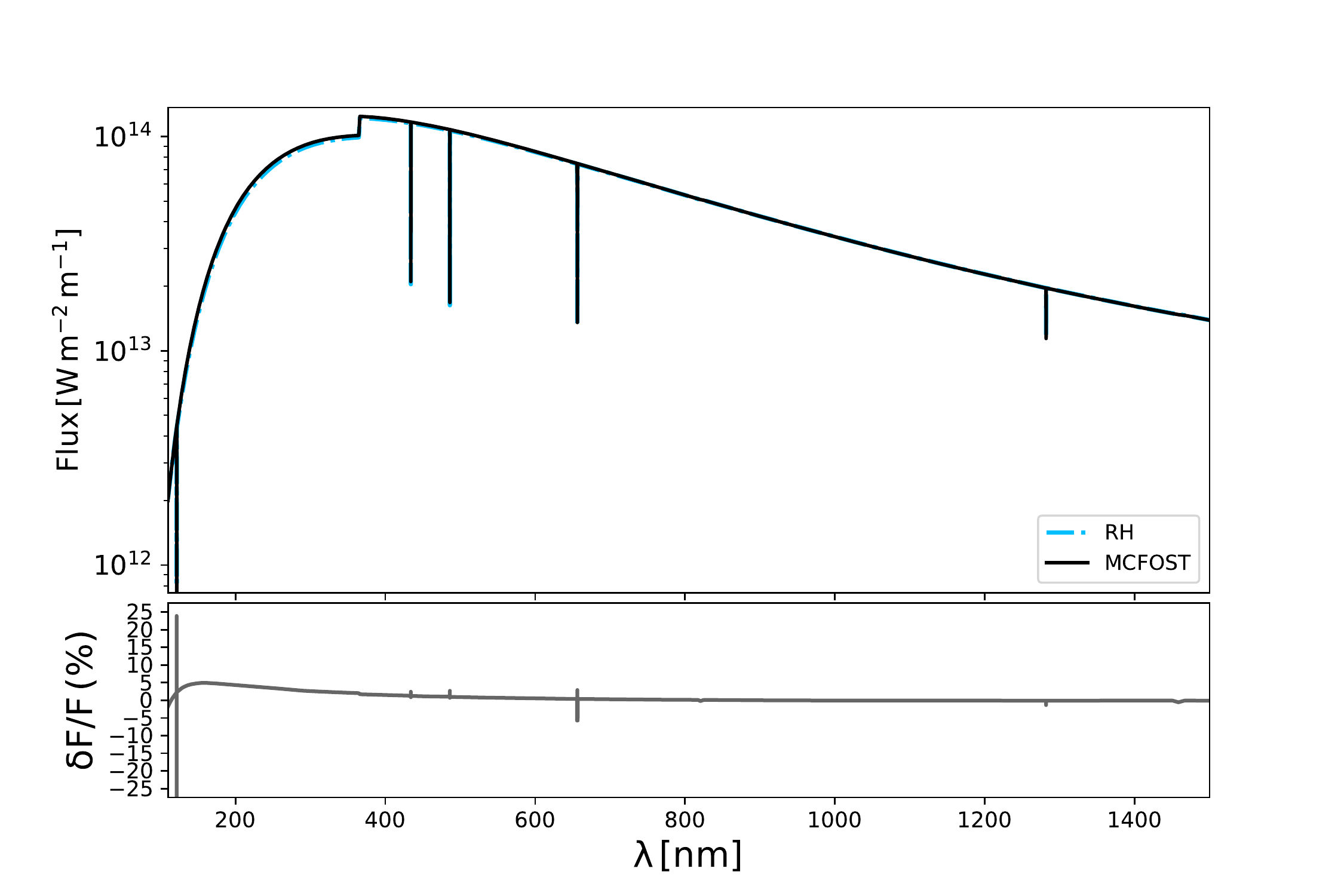}
 \caption{Non-LTE flux.}
\end{subfigure}

\begin{subfigure}{1\columnwidth}
  \centering

  \includegraphics[width=1\columnwidth]{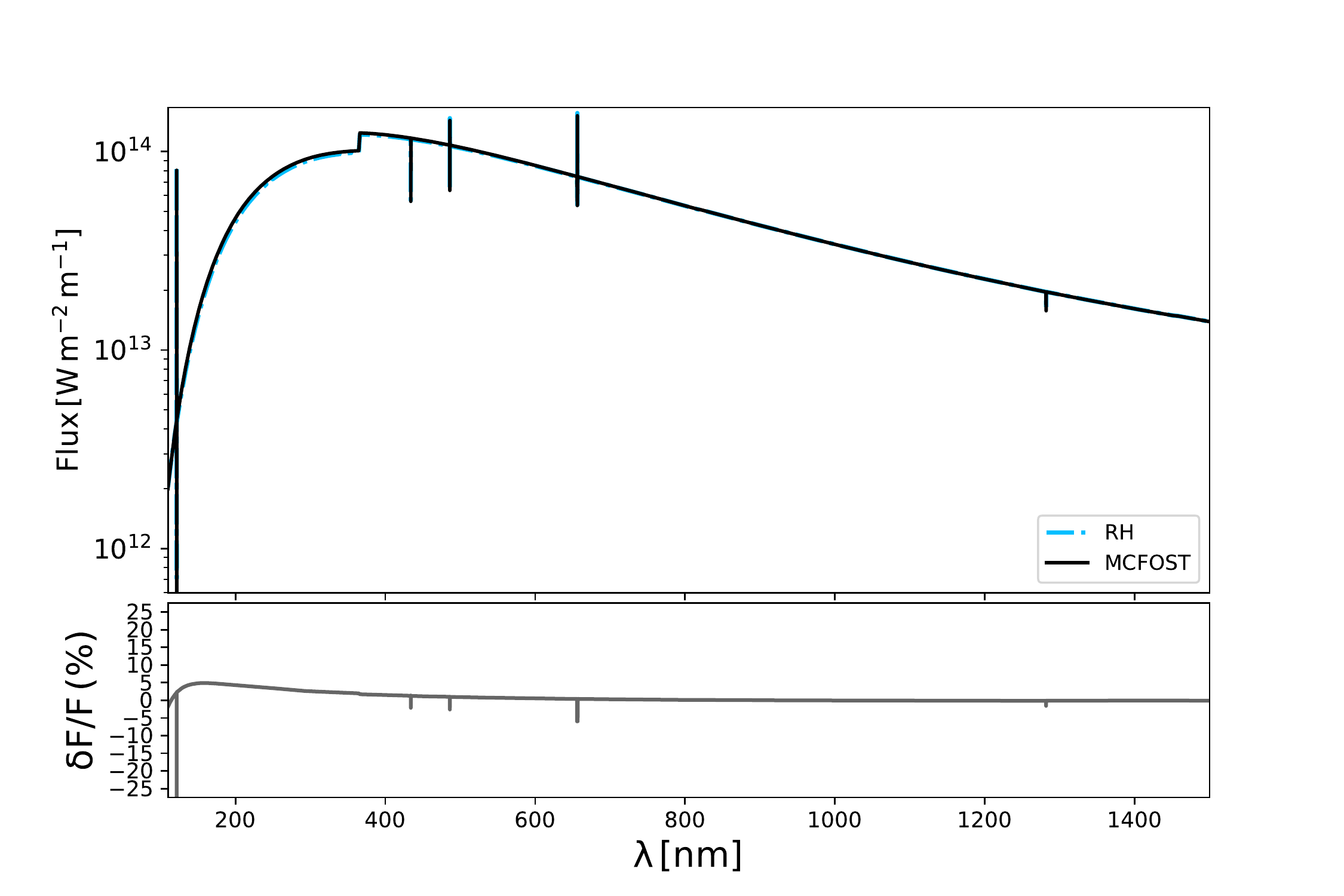}
  
 \caption{LTE flux.}
\end{subfigure}

  \caption{Solar flux for the model C. 
  RH is shown with the thin cyan and MCFOST with the thick black line. Below each panel the discrepancy between the two codes is shown.}
    \label{fig:flux_falc}
    
\end{figure}

Figure \ref{fig:flux_falc} shows the solar flux computed in model C with RH and MCFOST, at non-LTE (a) and LTE (b). The impact of non-LTE effects on spectral lines is clearly visible. At LTE all lines are in emission, whereas at non-LTE most of the lines are seen in absorption. For comparison, disc-integrated observations of the Sun show that most of the hydrogen lines are in absorption, except for strong chromospheric lines (e.g. Ly$\alpha$).

\begin{figure}[H]

\begin{subfigure}{\columnwidth}
  \centering
   \includegraphics[width=\textwidth]{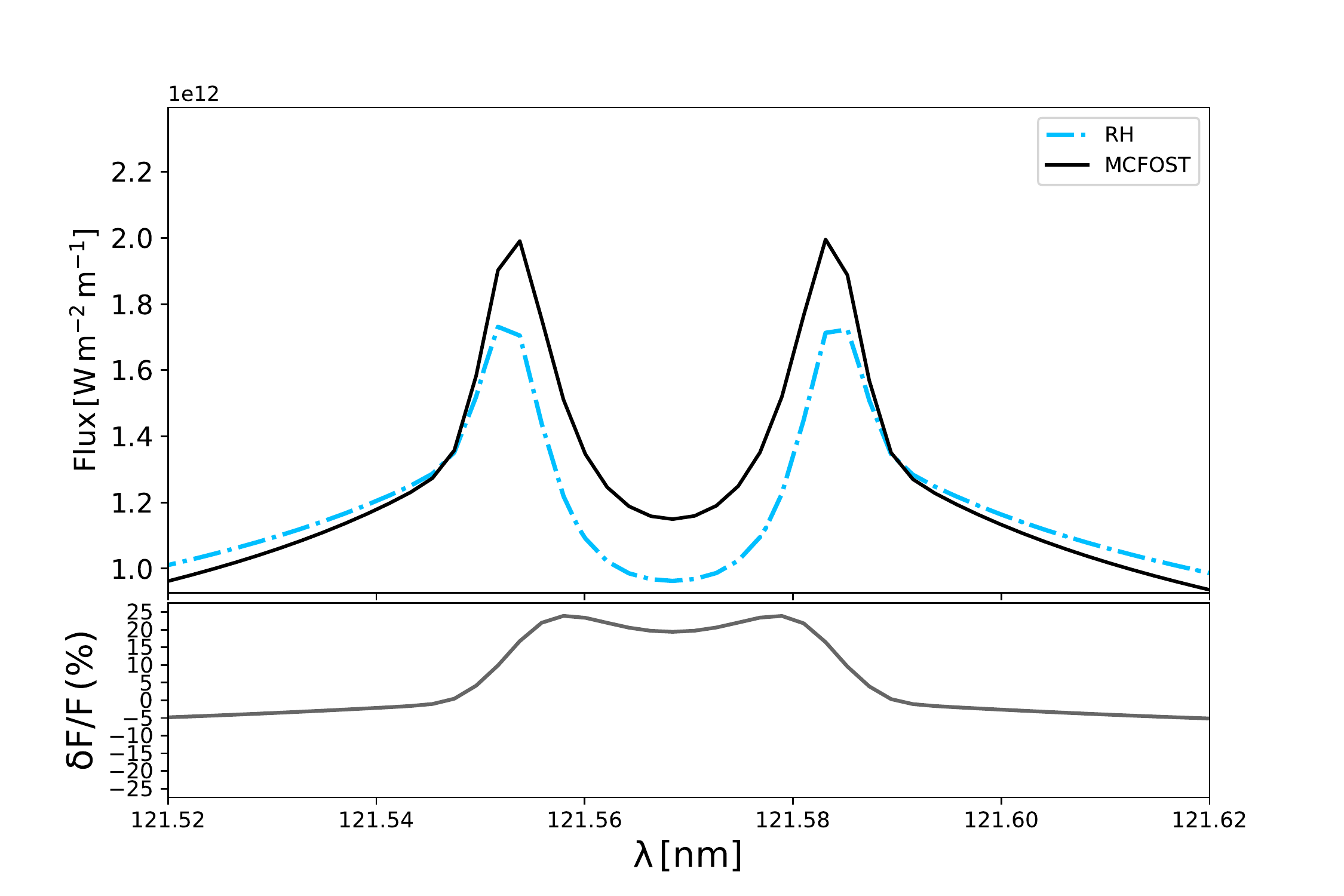}
 \caption{Ly$\alpha$ line.}
\end{subfigure}

\begin{subfigure}{\columnwidth}
  \centering

  \includegraphics[width=\textwidth]{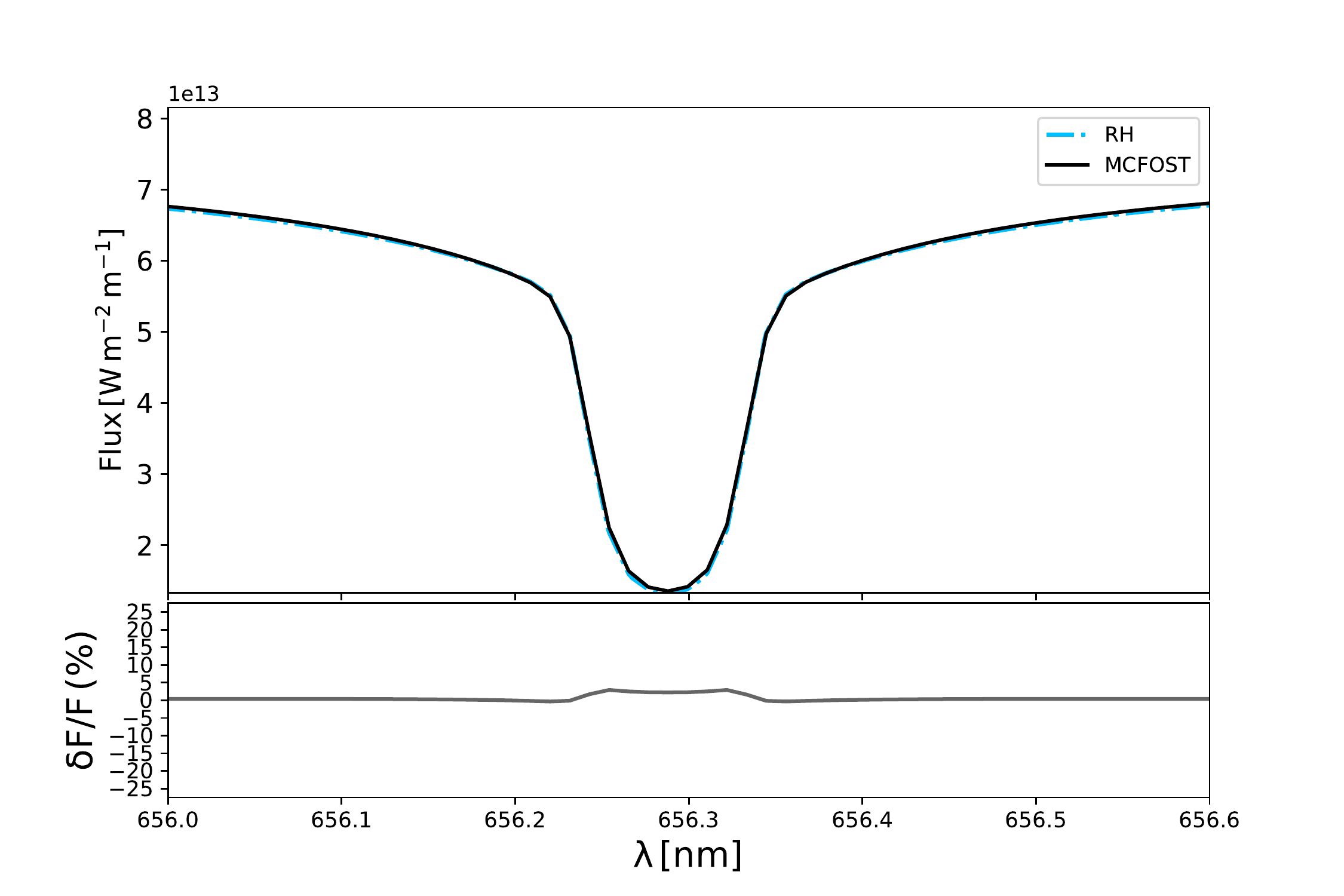}
  
 \caption{ \textsc{H}$\alpha$ line.}
\end{subfigure}

  \caption{Zoom-in on solar lines flux. The discrepancy between RH and MCFOST is shown below each panel.}
    \label{fig:flux_falc_lines}
    
\end{figure}

At all wavelengths, the discrepancy is below 1\%, except in the core of some lines where it reaches 5\%. In the figures, interpolation errors create glitches in the discrepancy functions.
However, real differences exist in the core of strong lines, such as the Ly$\alpha$ line, where the mismatch can reach 20\%-25\%. Figure \ref{fig:flux_falc_lines} shows the non-LTE flux of the Ly$\alpha$ line (top panel) and the flux of the \textsc{H}$\alpha$ line (bottom panel). 
Although these differences are not critical, using one of the two options discussed in \S \ref{angle_quad} to improve the angular quadrature leads to a better agreement between the Lyman $\alpha$ line computed by RH and MCFOST.

\subsection{Discussion on the angular quadrature}\label{accurate_pops}

\begin{figure}[H]

  \centering
   \includegraphics[width=1\columnwidth]{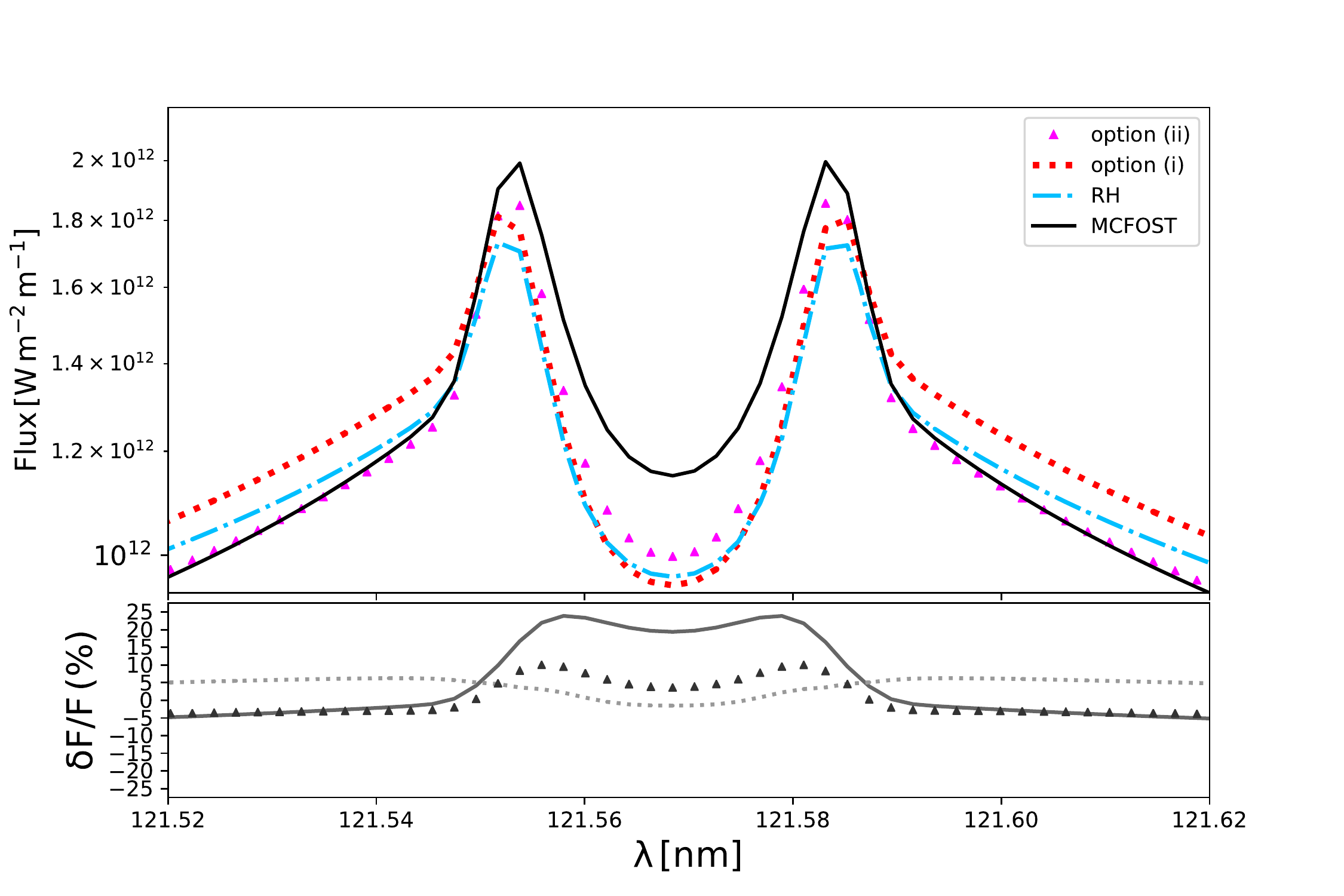}
  \caption{Evolution of Ly$\alpha$ flux with angular quadrature methods. The black and cyan lines are the flux computed by MCFOST and RH, respectively, from Fig. \ref{fig:flux_falc_lines}a. The dotted red line and the pink triangles are the flux computed using option (i) and (ii) of \S \ref{angle_quad}, respectively. The bottom panel shows the discrepancy between RH and MCFOST, from Fig. \ref{fig:flux_falc_lines}a (dark grey line), option (i) (light grey dotted line), and option (ii) (grey triangles).}
    \label{fig:flux_lyalpha_comp}
\end{figure}

In \S \ref{angle_quad}, we presented the method we used to perform angular quadratures. These quadratures are important since they are used to evaluate the radiative rates, which in turn determine the level populations. As stated earlier, angular quadratures are performed for a set of fixed rays (i.e. directions) starting from each cell centre. 
In the following, we test option (i) of \S \ref{angle_quad}, that is performing the angular quadrature not only from cell centres but also at N random positions in the cells. As in section \ref{solar_atm}, we compute the non-LTE Ly$\alpha$ line flux in model C. Further, we use the original grid of the 1D model, that is without additional points.
We randomly chose 99 additional positions in each cell. We then kept these positions fixed during the non-LTE loop to avoid introducing Monte Carlo noise in the solution. For each of these 100 positions, the angular quadrature was performed for 80 rays resulting in a total of 8\,000 rays. Thus, for each cell, we obtained 100 values of the mean intensity and radiative rates, each at a different location inside the cell. The mean intensity and the radiative rates in each cell is just the arithmetic mean of these values. \\
Finally, we interpolated the level populations on the same grid used in \S \ref{solar_atm} (i.e. with a higher resolution). 
Because now we have a better estimation of the radiative rates for each cell, the error in the numerical integration of Eq. \ref{eq:RTE} dominates the discrepancy between RH and MCFOST.
\\
In figure \ref{fig:flux_lyalpha_comp} we show the Ly$\alpha$ flux computed with this method compared to the profile shown in in Fig. \ref{fig:flux_falc_lines}a. The agreement between RH and MCFOST is substantially improved, with a maximum discrepancy below 5\%, five times better than with only one position for the angular quadrature. Still we note that, even if the core is better reproduced, the wings are not, although the maximum error is small.

Similar results are obtained by performing a purely Monte Carlo step after the one point quadrature, that is we randomly selected 1000 positions and directions (\S \ref{angle_quad}, option (ii)). The main difference between \S \ref{angle_quad} option (i) and \S \ref{angle_quad} option (ii) lies in the execution time of the code. The latter approach is eight times faster than the former for an equivalent accuracy. However, the populations computed with \S \ref{angle_quad} option (ii) have Monte Carlo noise.

\section{Conclusions}\label{conclusion}

In this paper, we tested our code against spherically symmetric models of stellar atmospheres, representing a range of physical plasma conditions. We tested background opacity calculations with lines merging close to series limits using the occupation probabilities formalism. The agreement between MCFOST-art and both TURBOspectrum and RH is excellent, of the order of 1\% or below at most wavelengths. Still, a few discrepancies are present, arising from the different treatment of opacities at a few critical wavelengths (e.g. at the Balmer jump or at the \text{H}$^-$ minimum). The code has also been successfully tested with vertical and rotational velocities. The proper behaviour of all spectral lines is recovered in models with non-zero velocity fields.

We performed non-LTE calculations of the solar Lyman $\alpha$ and H$\alpha$ lines formation. These calculations were compared with calculations performed with the RH code. 
Line formation regions and non-LTE effects are reproduced well by our code compared to earlier studies and observations. While the discrepancy in the flux between RH and MCFOST can reach 25\% in the core of Ly$\alpha$, the discrepancy drops below 1\% around the H$\alpha$ line and for most other wavelengths in the spectrum.
Furthermore, we showed that improving the accuracy of the angular quadrature scheme decreases this discrepancy by a factor of five.

Although we only tested the code against 1D models, the opacity modules and the non-LTE loop are geometry independent. These modules are therefore applicable to multidimensional models. The MCFOST code is 3D in essence and the solution of the radiative transfer equation is fully performed in 3D.
Further applications of this new code to 2D and 3D geometries will be discussed in a forthcoming paper with a particular focus on stellar magnetospheres.

\begin{acknowledgements}
We are thankful to Dr. Bertrand Plez for providing and maintaining a publicly available version of TURBOspectrum. B Tessore is grateful to Dr. Han Uitenbroek for providing a version of RH. We thank the anonymous referee for his comments, which helped to improve the manuscript.
B. Tessore is also thankful to the french minister of Europe and foreign affairs and the minister of superior education, research and innovation (MEAE and MESRI) for research funding through FASIC partnership.
This project has received funding from the European Research Council (ERC) under the European Union’s Horizon 2020 research and innovation programme (grant agreement No 742095; {\it SPIDI}: Star-Planets-Inner Disk-Interactions, \url{http://www.spidi-eu.org}). C. Pinte acknowledges funding from the Australian Research Council via FT170100040 and DP180104235.

\end{acknowledgements}

\bibliographystyle{aa}
\bibliography{bt_spidi1.bib}

\end{document}